\documentclass[twocolumn,astrosymb]{aastex7}

\usepackage{amsmath}
\usepackage{booktabs}

\newcommand{\kms}{\mathrm{km\ s^{-1}}}
\newcommand{\msun}{M_{\odot}}
\newcommand{\mbh}{M_{\mathrm{BH}}}
\newcommand{\mstar}{M_{\star}}
\newcommand{\ssfr}{\mathrm{sSFR}}
\newcommand{\ha}{\mathrm{H}\alpha}
\newcommand{\hb}{\mathrm{H}\beta}
\newcommand{\hdelta}{\mathrm{H}\delta}
\newcommand{\mgii}{\mathrm{Mg\,II}}
\newcommand{\oii}{[\mathrm{O\,II}]}
\newcommand{\oiii}{[\mathrm{O\,III}]}
\newcommand{\nii}{[\mathrm{N\,II}]}
\newcommand{\sii}{[\mathrm{S\,II}]}
\newcommand{\caii}{\mathrm{Ca\,II}}

\begin{document}

\title{Changing-Look AGNs from DESI. VI. Host Galaxies}

\author[orcid=0000-0002-1234-552X]{Shengxiu Sun}
\affiliation{Department of Astronomy, School of Physics, Peking University, Beijing 100871, China}
\affiliation{Kavli Institute for Astronomy and Astrophysics, Peking University, Beijing 100871, China}
\affiliation{Institute for Theoretical Physics, Heidelberg University, Philosophenweg 12, D-69120 Heidelberg, Germany}
\email{sxsun@stu.pku.edu.cn}

\author[orcid=0000-0003-4176-6486]{Linhua Jiang}
\affiliation{Department of Astronomy, School of Physics, Peking University, Beijing 100871, China}
\affiliation{Kavli Institute for Astronomy and Astrophysics, Peking University, Beijing 100871, China}
\email{jiangKIAA@pku.edu.cn}

\author[0000-0001-9457-0589]{Wei-Jian Guo}
\affiliation{Key Laboratory of Optical Astronomy, National Astronomical Observatories, Chinese Academy of Sciences, Beijing 100012, China}
\email{guowj@bao.ac.cn}

\author[orcid=0000-0001-8582-7012]{Sarah E.~I.~Bosman}
\affiliation{Institute for Theoretical Physics, Heidelberg University, Philosophenweg 12, D-69120 Heidelberg, Germany}
\affiliation{Max-Planck-Institut für Astronomie, Königstuhl 17, 69117 Heidelberg, Germany}
\email{bosman@thphys.uni-heidelberg.de}

\author[0000-0003-0230-6436]{Zhiwei Pan}
\affiliation{Department of Astronomy, University of Illinois Urbana-Champaign, Urbana, IL 61801, USA}
\email{zhiweip@illinois.edu}

\begin{abstract}

Changing-look (CL) AGNs trace rapid changes in nuclear activity, but their connection to host galaxy properties remains unclear. We present a study of the host galaxies of 105 CL AGNs previously selected by comparing DESI and SDSS data. We apply a two-epoch spectrophotometric decomposition to the DESI and SDSS spectra of the 105 objects. Meanwhile, HSC images are used to constrain their varying AGN components and non-varying stellar population components. We find that 79 of the 105 (75.2\%) CL AGN hosts are quiescent galaxies, and 31/105 (29.5\%) also show post-starburst signatures. 
We focus on 82 CL AGNs with extended host emission in the HSC images and compare them with extended quasars at similar redshift and stellar mass. Their star formation activity, Balmer absorption, and quiescent fractions are broadly consistent with those of the comparison quasars, although post-starburst hosts are more common among the CL AGNs. Our CL AGNs with extended host emission are more often quiescent than those with compact morphology, but this difference is not apparent after matching in redshift and stellar mass.
The $\oii$ and $\oiii$ narrow lines show no population-wide response to the continuum and broad line changes, consistent with the slower response expected from the narrow line region. Together, these results favor changes in the central supermassive black hole accretion rate as the main origin of the CL transitions.

\end{abstract}


\section{Introduction}\label{sec:intro}

Changing-look (CL) AGNs are AGNs whose broad UV/optical emission lines appear, disappear, or change substantially in different  epochs, usually with large continuum variations. These transitions occur on timescales of months to years, posing a challenge to the AGN unification model that AGN types are primarily determined by viewing angles \citep[e.g.,][]{LaMassa2015,MacLeod2016}. The main proposed explanations are changes in the central accretion flow, changes in line-of-sight obscuration, and tidal disruption events (TDEs) or TDE-like flares. In the first case, a change in the accretion rate or inner disk structure modifies the ionizing continuum and can drive correlated responses from the broad line region (BLR) and hot dust \citep{Sheng2017MIR,NodaDone2018,Ross2018CLQ,Stern2018MIRCLQ}. In the second case, variable obscuration changes the observed continuum and broad line strength without requiring the central engine itself to fade \citep{Goodrich1995Dust,Trippe2010Seyfert}. In the third case, a transient flare powered by stellar debris can temporarily enhance the nuclear continuum and excite gas in the AGN environment \citep{Merloni2015TDE,Blanchard_2017}. Distinguishing among these possibilities requires repeated spectroscopy together with optical, infrared, and, where available, X-ray constraints.

Large spectroscopic surveys have shifted CL AGN studies from individual objects to the population level. Studies based on the comparison between spectra from the Dark Energy Spectroscopic Instrument (DESI) Data Release 1 (DR1) and the Sloan Digital Sky Survey (SDSS) identified 561 CL AGNs at $z\leq0.9$ \citep{Guo2025DR1}. For most objects in this parent sample, their broad line and optical continuum changes, together with the mid-infrared behavior, favored variations in the ionizing continuum over simple foreground extinction \citep{2024ApJS..270...26G_CLAGN_from_EDR,Guo2025DR1}. A subsequent analysis of the broad emission lines found an ordered sequence of line response to the fading continuum \citep{Guo2025BLR}. These results motivate a complementary question whether the host galaxies of CL AGNs differ from those of other quasars at similar redshift and stellar mass.

Host galaxies provide information on the longer timescale environments of CL AGNs. If CL episodes are preferentially associated with galaxy transformation, their hosts might show excesses of recently quenched stellar populations, disturbed morphologies, or offsets from black hole scaling relations with host galaxy properties. If their host properties are similar to those of non-CL quasar hosts, the changing-look phenomenon is more likely to be governed mainly by processes in the nucleus. Existing studies of CL AGN host galaxies have not reached a conclusion. \citet{2021ApJ...907L..21D} studied 15 CL AGNs and found that 8 resided in the green valley; their hosts also had centrally concentrated light profiles and showed little evidence for recent major mergers. From spectral analysis of 26 turn-off CL AGNs, \citet{Jin2022} found that their stellar populations were broadly similar to those of non-CL AGNs but had a larger contribution from stars of intermediate age. In contrast, all five CL AGNs studied with Mapping Nearby Galaxies at Apache Point Observatory (MaNGA) lay on the star-forming main sequence and had host properties broadly similar to non-CL MaNGA AGNs \citep{Yu2020}. A study of 82 turn-on CL AGNs found that their hosts followed local black hole scaling relations and had star formation rates (SFRs) comparable to type~2 AGN hosts \citep{Yang2025}.

Recent work has emphasized the need for explicit comparison samples. \citet{Verrico2025} compared 39 CL AGNs with Seyfert galaxies selected to have similar basic properties and found no significant difference in their recent star formation histories (SFHs), nor evidence for rapid quenching across galaxies. \citet{Zeltyn2026} analyzed spectra at intermediate resolution for 23 CL AGNs from SDSS-V and found host stellar populations consistent with those of type~2 AGNs. In contrast, \citet{Tian2026Morphology} visually inspected 63 CL AGN hosts at $z<0.15$ and reported a higher merger fraction than those in their comparison samples. The different redshift ranges, comparison samples, and host measurements limit direct comparisons among these studies.

Separating the AGN and host galaxy components is the main challenge for host galaxy studies of CL AGNs. In spectral fitting alone, the AGN continuum and stellar contribution are degenerate, particularly when the central engine is bright. Imaging constrains the spatially resolved host contribution and reduces this degeneracy. In \citet{Sun2026}, we developed a spectrophotometric decomposition method that combines DESI spectroscopy with five-band Subaru Hyper Suprime-Cam (HSC) imaging for 1083 quasars with extended morphology. The analysis found a predominantly quiescent host population and a 23\% post-starburst fraction. For CL AGNs, repeated spectra offer an additional consistency test because the stellar population should remain unchanged while the nuclear component varies \citep{Aydar2026}.

In this work, we adapt the method of \citet{Sun2026} for two spectral epochs CL AGNs. We assume that the same host galaxy stellar population underlies the DESI and SDSS spectra, while allowing the AGN emission and the host contribution within each aperture to vary between epochs. The paper is organized as follows. Section~\ref{sec:data} describes the parent CL AGN sample, HSC imaging data, morphology classification, and light curves from multiple surveys. Section~\ref{sec:method} presents the two-epoch spectrophotometric decomposition and the matched comparison sample of extended quasars. Section~\ref{sec:results} reports the host stellar population measurements and results as a function of morphology. Section~\ref{sec:discussion} discusses the variability on long timescales, forbidden line stability, black hole scaling relations, and implications for the changing-look mechanism. Section~\ref{sec:summary} summarizes the main conclusions. We adopt a flat $\Lambda$ cold dark matter ($\Lambda$CDM) cosmology with $H_0=70\ \kms\ \mathrm{Mpc}^{-1}$, $\Omega_m=0.3$, and $\Omega_\Lambda=0.7$.

\section{Data and Sample}\label{sec:data}

\subsection{Parent CL AGN Sample}

Our parent sample is the DESI DR1 CL AGN sample of \citet{Guo2025DR1}. It was constructed by comparing DESI DR1 and archival SDSS DR16 spectra \citep{2020ApJS..249....3A}, measuring changes in broad $\ha$, $\hb$, $\mgii$, and ultraviolet emission lines where available, and confirming the spectral transitions with the aid of optical light curves. The catalog contains 561 objects at $z\leq0.9$, including 283 turn-on and 278 turn-off CL AGNs. We adopt the state definition of the parent catalog: ``turn-on'' means that the DESI epoch has the stronger broad-line and AGN continuum emission, while ``turn-off'' means that the earlier SDSS epoch is brighter.

DESI spectra cover approximately $3600$--$9800$ \AA{} at a resolving power of $R\sim2000$--5000 through 1\farcs5-diameter fibers \citep{DESI2016a.Science,DESI2016b.Instr,DESI2022.KP1.Instr,FocalPlane.Silber.2023,Spectro.Pipeline.Guy.2023,SurveyOps.Schlafly.2023,2025arXiv250314745D}. The archival SDSS spectra span several phases: SDSS-I/II Legacy Survey spectroscopy used 3\arcsec-diameter fibers, whereas the Baryon Oscillation Spectroscopic Survey (BOSS) and extended Baryon Oscillation Spectroscopic Survey (eBOSS) used 2\arcsec-diameter fibers \citep{2000AJ....120.1579Y,2013AJ....146...32S}. In the spectrophotometric decomposition and the HSC aperture photometry, we use the SDSS fiber diameter associated with each spectrum observation. 

\subsection{HSC Imaging Data and the Final CL AGN Sample}

\begin{figure*}[t]
\centering
\includegraphics[width=\textwidth]{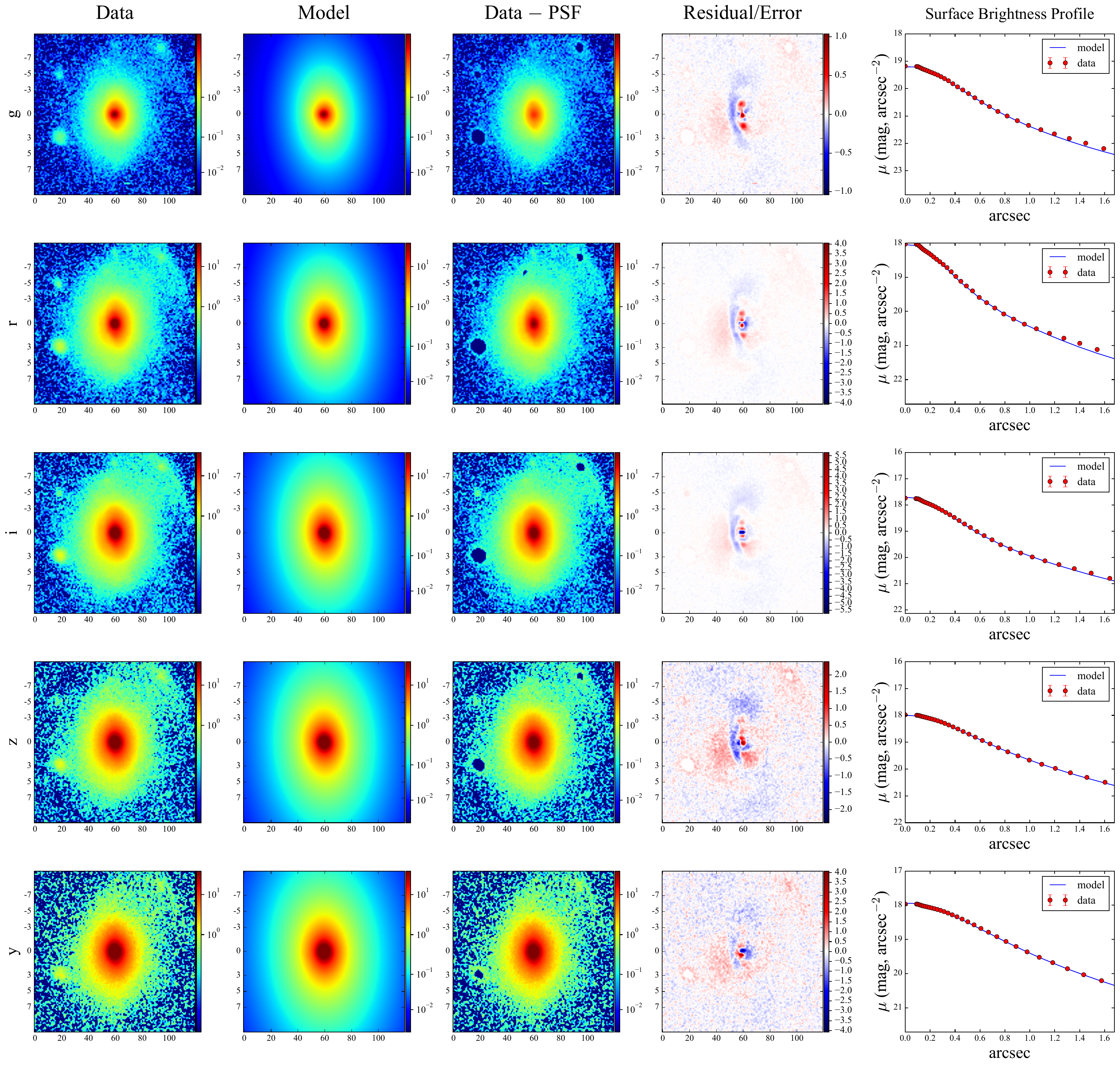}
\caption{Example five-band HSC imaging decomposition for internal identification number (ID) 121. From left to right, the columns show the image, model, point-source-subtracted image, normalized residual, and surface-brightness profile from the \texttt{GalfitM} fitting result.}
\label{fig:galfitm_example}
\end{figure*}

We crossmatch the parent catalog with the $g,r,i,z,y$ imaging in the third data release of the Hyper Suprime-Cam Subaru Strategic Program \citep{Miyazaki2018HSC,HSC_DR3_2022}. The match yields 137 CL AGNs. For each source, we fit the five bands simultaneously with \texttt{GalfitM} using a point spread function (PSF) component for the AGN and a single S\'ersic component for the host galaxy \citep{2022A&A...664A..92H_galfitm}. Nearby sources are masked, and PSF models for each CL AGN are generated using the HSC software pipeline \texttt{hscPipe} v8 \citep{HSC_pipeline,LSST_pipe_2019ASPC..523..521B}. We measure the host flux directly in the PSF-subtracted images with aperture photometry matching the DESI and SDSS fiber diameters. The resulting host galaxy flux levels provide constraints on the host galaxy component in the subsequent spectrophotometric decomposition \citep[see detailed descriptions in][]{Sun2026}.
Figure~\ref{fig:galfitm_example} shows one representative HSC imaging decomposition.

Starting from the 137 HSC-matched objects, two-epoch spectrophotometric decomposition results are available for 130 objects; the other 7 are excluded due to low signal to noise ratios (S/Ns). Applying the fitting quality criteria that require reduced $\chi^2\leq3$ and a physical $\hdelta$ absorption strength $-10<\hdelta_A/\text{\AA}<20$ to avoid abnormal decomposition, we exclude 20 objects that do not satisfy all requirements for the host property analysis. The main reason is that their host continuum emission is too weak. 
The HSC imaging quality assessment described in Section~\ref{sec:morph_method} excludes five additional objects from the remaining 110 objects. The final sample thus contains 105 CL AGNs, including 56 turn-off and 49 turn-on systems. The morphology classification in Section~\ref{sec:morph_method} identifies extended host galaxy emission in 88 objects, while 17 are compact at the HSC resolution. Of the 88 objects with extended host emission, 83 have at least one comparison quasar satisfying the matching criteria in Section~\ref{sec:matching}. Table~\ref{tab:final_sample} lists the properties of the 105 objects and the internal IDs used throughout this paper. The multi-survey light curves are compiled for all 137 HSC-matched objects, including sources that do not enter the final host galaxy sample.

\begin{deluxetable*}{r l c c c c c c c c c}
\tabletypesize{\scriptsize}
\tablecaption{Final CL AGN host galaxy sample\label{tab:final_sample}}
\tablehead{
\colhead{ID} & \colhead{SDSS Name} & \colhead{R.A.} & \colhead{Decl.} & \colhead{$z$} & \colhead{State} & \colhead{$g_{\mathrm{tot}}$} & \colhead{$g_{\mathrm{host}}$} & \colhead{$\log L_{5100,\mathrm{on}}$} & \colhead{$\log M_\star$} & \colhead{Host morphology}\\
 & & \colhead{(deg)} & \colhead{(deg)} & & & \colhead{(AB mag)} & \colhead{(AB mag)} & \colhead{(erg s$^{-1}$)} & \colhead{($M_\odot$)} & 
}
\startdata
0 & J000048.17+013313.6 & 0.20072 & 1.55379 & 0.6640 & turn-off & 19.31 & 20.49 & 44.74 & 11.10 & compact \\
23 & J011311.82+013542.4 & 18.29925 & 1.59513 & 0.2375 & turn-on & 20.50 & 20.93 & 43.80 & 10.56 & extended \\
28 & J011942.12-001901.9 & 19.92552 & -0.31721 & 0.2561 & turn-on & 21.70 & 21.94 & 43.27 & 10.30 & extended \\
30 & J011958.08+004350.2 & 19.99201 & 0.73064 & 0.7000 & turn-off & 22.19 & 23.09 & 44.16 & 10.44 & compact \\
32 & J012217.42-001759.2 & 20.57261 & -0.29980 & 0.3230 & turn-off & 20.35 & 21.99 & 43.85 & 10.87 & extended \\
33 & J012353.46+000938.8 & 20.97280 & 0.16078 & 0.6543 & turn-on & 22.75 & 24.15 & 43.90 & 10.74 & compact \\
48 & J015813.43-014432.0 & 29.55599 & -1.74221 & 0.5011 & turn-on & 22.19 & 22.86 & 43.52 & 10.95 & compact \\
50 & J015845.78-002138.5 & 29.69077 & -0.36071 & 0.5640 & turn-off & 21.91 & 23.39 & 44.28 & 10.94 & extended \\
51 & J020024.11+000535.6 & 30.10050 & 0.09321 & 0.3936 & turn-off & 21.15 & 21.53 & 43.84 & 10.70 & extended \\
53 & J020457.53-005755.7 & 31.23971 & -0.96549 & 0.4450 & turn-off & 21.89 & 22.31 & 43.87 & 11.01 & extended \\
54 & J020517.22-061431.6 & 31.32179 & -6.24215 & 0.7180 & turn-off & 21.93 & 24.32 & 43.80 & 10.72 & compact \\
56 & J020744.06-060955.9 & 31.93359 & -6.16556 & 0.6490 & turn-off & 20.21 & 23.48 & 44.48 & 10.63 & extended \\
\enddata
\tablecomments{The table lists the first twelve rows of the final 105-object host galaxy sample; the complete table is available in machine-readable form. The first column gives the internal decomposition index used throughout this analysis. The second column gives the SDSS source name, and the third and fourth columns give the right ascension and declination measured by DESI. The sixth column follows the DESI--SDSS state definition from the parent CL AGN catalog. The two $g$-band columns are HSC AB magnitudes measured from the final \texttt{GalfitM} decomposition within a 1\farcs5-diameter aperture matched to the DESI fiber: $g_{\mathrm{tot}}$ is the sum of the point-source and host components, and $g_{\mathrm{host}}$ is the host-only component. The corresponding SDSS-aperture host fluxes, measured with the fiber diameter of each SDSS spectrum, are used in the two-epoch decomposition but are not listed here. The ninth column gives the on-state AGN component $\lambda L_{\lambda}(5100\,\text{\AA})$ measured from the decomposition. Host morphology is listed as extended or compact according to the HSC imaging decomposition.}
\end{deluxetable*}

\subsection{Host Galaxy Morphology}\label{sec:morph_method}

We use the \texttt{GalfitM} fitting results to identify CL AGNs with extended host morphology.
For each object, we measure the intrinsic $i$-band S\'ersic effective radius $R_{e,i}$ and the half-light radius of its fitted HSC PSF, $R_{50,\mathrm{PSF},i}$. We classify sources with $R_{e,i}/R_{50,\mathrm{PSF},i}>1$ as extended morphology and those with $R_{e,i}/R_{50,\mathrm{PSF},i}\leq1$ as compact morphology. This compact label is observational and reflects the limits of the HSC decomposition rather than a physical size classification, because surface brightness dimming and PSF--S\'ersic profile degeneracy vary with redshift and data quality. The morphology classifications for the final sample are listed in Table~\ref{tab:final_sample}.

For each source, we visually inspect the PSF-subtracted images, normalized residuals, and surface brightness profiles from the \texttt{GalfitM} fitting results. We exclude fits when the fitting diagnostics failed, the reduced imaging-fit $\chi^2$ exceeds 20, $R_{e,i}>60$ pixels (10\farcs1; half the cutout width), the S\'ersic index reaches the fitting boundary at $n_i>19.5$, or the PSF--host centroid offset exceeds 5 pixels (0\farcs84). We identify ten systems whose PSF-subtracted images show asymmetry, tidal features, or multicomponent structure; eight of these systems remain in the final sample.

\subsection{Optical and Mid-Infrared Light Curves}\label{sec:lightcurve_method}

Following the DESI CL AGN light curve analyses of \citet{2024ApJS..270...26G_CLAGN_from_EDR,Guo2025DR1}, we compile optical light curves for the HSC-matched sources from the Panoramic Survey Telescope and Rapid Response System 1 (Pan-STARRS1; PS1) \citep{Chambers2016PS1,Flewelling2020}, Palomar Transient Factory (PTF) \citep{Law2009PTF,Ofek2012PTF}, Zwicky Transient Facility (ZTF) \citep{Masci2019}, and Catalina Real-Time Transient Survey (CRTS) when retrievable \citep{Drake2009}. We compile the mid-infrared light curves from the Wide-field Infrared Survey Explorer (WISE) and Near-Earth Object Wide-field Infrared Survey Explorer Reactivation Mission (NEOWISE) \citep{Wright2010,Mainzer2014}. We use the nearest catalog source within 3\arcsec\ of the DESI position, apply quality cuts similar to those used in the DESI CL AGN studies, and bin the optical and mid-infrared measurements over 10 and 180 days, respectively. We adopt the calibration convention of each survey and analyze differential magnitudes relative to the median of each survey and filter series. The complete light curve atlas is provided as online supplementary material.

When spectrophotometry from the Spectro-Photometer for the History of the Universe, Epoch of Reionization, and Ices Explorer (SPHEREx) Quick Release 2 (QR2) is available, we treat each observed spectrum as one infrared epoch \citep{Bock2026SPHEREx,Akeson2025SPHEREx}. We synthesize WISE-like W1 and W2 Vega magnitudes by integrating the SPHEREx spectrum over the covered portion of the corresponding WISE response curve and include the measurements in the mid-infrared light curve where the wavelength coverage permits. We also convolve the SDSS and DESI spectra with the SDSS $g$ and $r$ response curves and mark the resulting synthetic AB magnitudes at the spectral epochs; these points are used only to mark the spectral epochs because of the SDSS and DESI fiber difference. The light curves confirm the CL behavior and identify recurrence or broad mid-infrared excursions relevant to TDE-like alternatives.

\section{Spectrophotometric Decomposition}\label{sec:method}

Our analysis is based on the spectrophotometric decomposition method developed by \citet{Sun2026} for DESI quasars with extended host morphology in the HSC images. In that work, the HSC imaging was modeled with an unresolved nuclear component and an extended host component, and the resulting five-band photometry of the host galaxy was used as an external constraint on the stellar contribution to the spectrum. This imaging constraint fixes the host galaxy flux scale independently of the spectral continuum fit and reduces the degeneracy between the featureless AGN continuum and the host stellar continuum. The decomposition is then iterated between the stellar and AGN components until the solution converges. Here we extend this framework from a single DESI spectrum to paired SDSS and DESI spectra of CL AGNs, allowing the nuclear emission to vary between epochs while explicitly treating the aperture difference between the two surveys.

\subsection{Two-Epoch Model}

Our model extends the single-epoch decomposition to two spectral epochs. In schematic form, the observed flux density at epoch $i$ is
\begin{equation}
F_{\lambda}^{\,i}=A_{\lambda}^{\,i}(\boldsymbol{\phi}^{\,i})
        +T^{\,i}\!\left[G_{\lambda}(\boldsymbol{\theta}_{\star})\right]
        +\epsilon_{\lambda}^{\,i},
\label{eq:model}
\end{equation}
where $A_{\lambda}^{\,i}$ is the AGN model for epoch $i$, $G_{\lambda}$ is the intrinsic host stellar population model, $T^{\,i}$ maps this stellar population into the fiber aperture of epoch $i$, and $\epsilon_{\lambda}^{\,i}$ is the residual. The stellar population parameters $\boldsymbol{\theta}_{\star}$ are shared by the DESI and SDSS epochs. In contrast, the AGN parameters $\boldsymbol{\phi}^{\,i}$, including continuum slopes and emission line strengths, are independent. In our implementation, $T^{\,i}$ is constrained by the HSC host photometry instead of simply applying a fiber size scaling. We use the host spectrum corresponding to the DESI aperture as the fiducial host galaxy component. The HSC imaging decomposition provides $g,r,i,z,y$ host galaxy photometry measured within the DESI aperture and within the SDSS aperture matched to the corresponding spectrum. The host photometry in the DESI aperture is first fit with a nonparametric SFH model for the spectral energy distribution (SED) to obtain the initial estimate of the fiducial host spectrum. The photometric difference between the SDSS and DESI apertures is then fit with the same SED model to estimate the additional host spectrum entering the SDSS aperture.

During each iteration, the DESI spectrum after AGN subtraction and the SDSS spectrum after subtracting both the AGN model and the aperture difference host component provide the two host spectra to be modeled. These two corrected host spectra are averaged and fit together with the DESI aperture host photometry to update the fiducial host galaxy model. The SDSS host component model is constructed by adding the aperture difference host spectrum back to this fiducial model. This treatment uses both the morphology information from HSC imaging and the two spectral epochs to constrain the host flux level, and is more flexible than a scalar aperture correction.

\begin{figure}[t]
\centering
\includegraphics[width=\columnwidth]{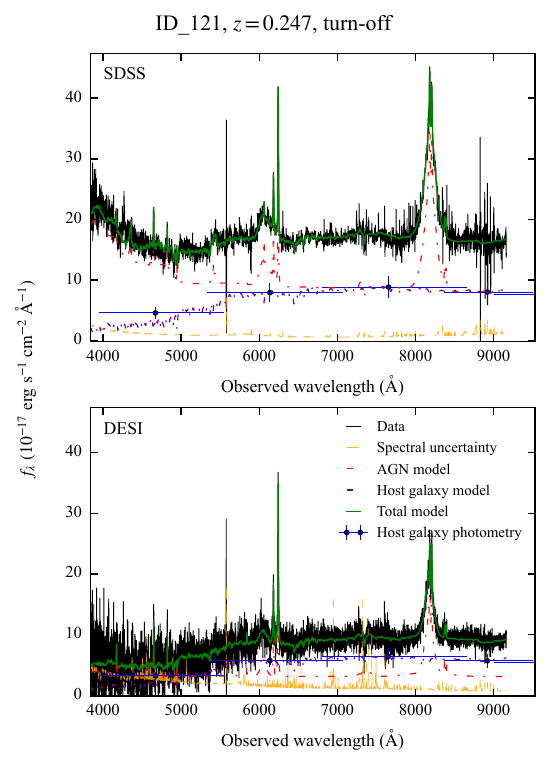}
\caption{Two-epoch spectrophotometric decomposition of ID 121, a representative turn-off CL AGN in our sample at $z=0.247$. The earlier SDSS spectrum is shown in the upper panel and the DESI spectrum in the lower panel. The model allows the AGN continuum and broad $\ha$ and $\hb$ emission to vary independently while tying the underlying stellar population across the two epochs.}
\label{fig:decomp_off}
\end{figure}

\begin{figure}[t]
\centering
\includegraphics[width=\columnwidth]{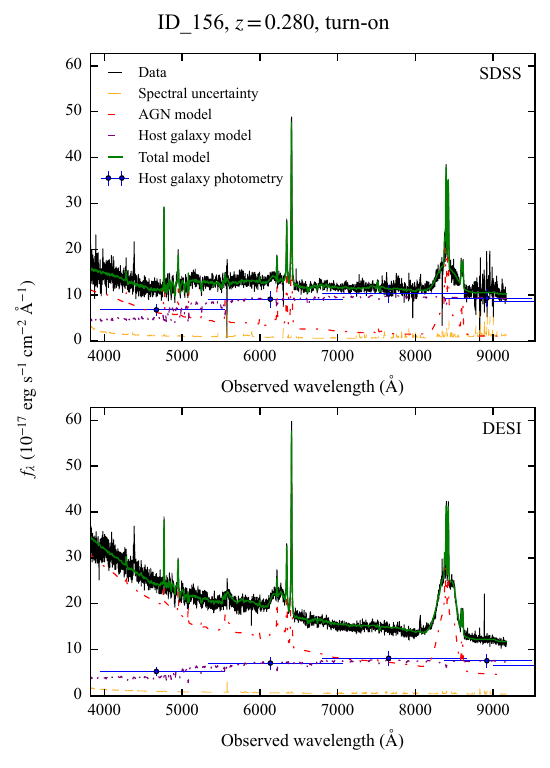}
\caption{Same as Figure~\ref{fig:decomp_off}, but for ID 156, a representative turn-on CL AGN in our sample at $z=0.280$. The host galaxy stellar absorption features are well modeled even when AGN emission dominates at the DESI epoch.}
\label{fig:decomp_on}
\end{figure}

\begin{figure}[t]
\centering
\includegraphics[width=\columnwidth]{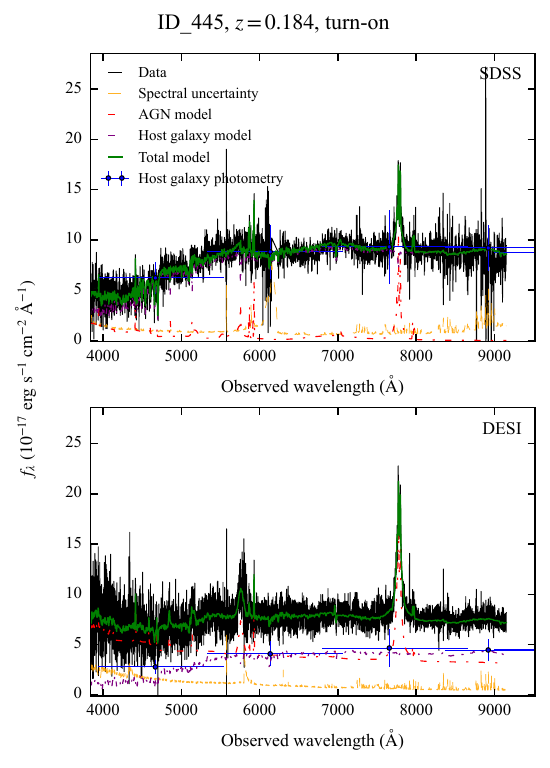}
\caption{Two-epoch spectrophotometric decomposition of ID 445, a turn-on CL AGN with an SDSS faint state nearly dominated by the host galaxy. The SDSS spectrum and HSC morphology information jointly anchor the host flux level.}
\label{fig:decomp_lowagn}
\end{figure}

Figures~\ref{fig:decomp_off} and~\ref{fig:decomp_on} present representative turn-off and turn-on CL AGN decomposition results, respectively. The changing AGN continuum and broad lines are captured independently at the two epochs, while the host model reproduces stellar absorption features such as the Balmer series and $\caii$ H and K. Figure~\ref{fig:decomp_lowagn} shows an additional case in which the faint state is nearly host-dominated.

\subsection{AGN and Host Stellar Components}

We use the same AGN continuum templates, fitting windows, and emission line parameterization as \citet{Sun2026}; further implementation details are provided there. The continuum model contains a broken power law, ultraviolet and optical Fe~II pseudocontinua, the Balmer continuum, and higher-order Balmer lines. The ultraviolet Fe~II template combines the templates of \citet{2001ApJS..134....1V}, \citet{Salviander2007}, and \citet{Tsuzuki_2006} over the rest-frame wavelengths of $1200$--$2200$, $2200$--$3090$, and $3090$--$3500$ \AA, respectively. The optical Fe~II template over $3686$--$7484$ \AA\ is from \citet{1992ApJS...80..109B}. The Balmer continuum and higher-order Balmer lines are modeled following \citet{Hu_2008}.
After subtraction of the fitted continuum, the emission lines are modeled with the \texttt{PyQSOFit} Gaussian parameterization \citep{2018ascl.soft09008G}. Broad $\ha$, $\hb$, and $\mgii$ are each represented by two Gaussians, with one additional Gaussian for the corresponding narrow component. Broad and narrow $\mathrm{H}\gamma$ and $\hdelta$ are each represented by one Gaussian. The two lines of the $\oiii\ \lambda\lambda4959,5007$ doublet each contain a core and a wing Gaussian, whereas $\oii$, $\nii$, $\sii$, and the other narrow lines are fitted with one Gaussian. The velocity offsets and widths are tied among narrow lines assigned to the same kinematic component. Broad and narrow components are separated at a full width at half maximum (FWHM) of $1200\ \kms$. Lines are fitted only when covered by the observed spectrum, and all AGN continuum and emission line parameters are allowed to vary independently between the two epochs.

The host component is fitted with \texttt{Bagpipes} \citep{10.1093/mnras/sty2169}. We use a nonparametric SFH with 14 age bins extending from the observation epoch to the age of the Universe and a Student-$t$ continuity prior following the approach of \citet{Leja_2019}. The fit returns $\mstar$, SFR, dust attenuation, stellar metallicity, and a luminosity-weighted stellar velocity dispersion. 

The fitting alternates between the AGN and host components. Starting from the stellar SED constrained by HSC, we subtract the host estimate and fit the residual AGN spectrum at each epoch. We then subtract the AGN models and jointly refit the two host spectra and their aperture photometry. Iteration stops when the principal host and AGN quantities change by less than 5\% between successive cycles. Sharing $\boldsymbol{\theta}_{\star}$ allows the epoch with higher host galaxy fraction to better constrain the host galaxy component.

\subsection{Stellar-Population Diagnostics}\label{sec:hdelta_method}

The specific star formation rate (sSFR) is defined as $\ssfr\equiv\mathrm{SFR}/\mstar$. Following the criteria adopted by \citet{2021ApJ...907L..21D} and \citet{Sun2026}, a host galaxy is classified as quiescent when
\begin{equation}
\log(\ssfr/\mathrm{yr}^{-1})<-10.94.
\end{equation}
We measure $\hdelta$ absorption strength $\hdelta_A$ directly from the decomposed DESI host spectrum using the Lick-index definition of \citet{1997ApJS..111..377W}.
A post-starburst host galaxy is defined as quiescent and having $\hdelta_A>4$ \AA. Measuring the index on the AGN-subtracted spectra reduces dependence on the adopted SFH prior.

\subsection{Black Hole Masses}\label{sec:Mbh}
Black hole masses are estimated from the decomposed AGN spectrum in the on state, using the same single-epoch virial calibrations as in \citet{Sun2026}. We use the on-state spectrum because the virial estimator becomes unstable when the broad line is weak or absent. For objects with reliable broad $\hb$ measurements, we use the \citet{2006ApJ...641..689V_VP06} calibration,
\begin{equation}
\begin{aligned}
\log\left(\frac{\mbh}{\msun}\right)
&=6.91
 +0.533\log\left[\frac{\lambda L_{\lambda}(5100\,\text{\AA})}
 {10^{44}\ \mathrm{erg\ s^{-1}}}\right] \\
&\quad +2\log\left[\frac{\mathrm{FWHM}(\hb)}
{10^3\ \kms}\right].
\end{aligned}
\end{equation}
For objects without a reasonable broad $\hb$ fit, we use the $\ha$ calibration of \citet{2005ApJ...630..122G},
\begin{equation}
\begin{aligned}
\frac{\mbh}{\msun}
&=2.0\times10^6
\left[\frac{L(\ha)}{10^{42}\ \mathrm{erg\ s^{-1}}}\right]^{0.55} \\
&\quad \times
\left[\frac{\mathrm{FWHM}(\ha)}{10^3\ \kms}\right]^{2.06}.
\end{aligned}
\end{equation}
The $\mgii$ line is fitted and included in the line variation measurements when it lies in the spectral coverage, but it is not used for the fiducial black hole masses in this sample at low redshift. A separate $\mgii$ mass calibration would be needed for objects without Balmer coverage \citep{2009ApJ...707.1334W}; recent DESI work also discusses Fe~II and effects associated with high accretion rates in masses based on $\mgii$ \citep{Mgii_MBH_pzw}. We do not apply those $\mgii$ estimates here because the scaling relation sample is mainly constrained by Balmer line masses and contains few objects for which a high accretion $\mgii$ correction would be relevant. At each epoch, we estimate the isotropic bolometric luminosity from the decomposed AGN continuum as $L_{\rm bol}^{\,i}=9.26\,\lambda L_{\lambda}^{\,i}(5100\,\text{\AA})$, using the fixed optical bolometric correction of \citet{Richards2006SED}. The Eddington ratio is $\lambda_{\rm Edd}^{\,i}=L_{\rm bol}^{\,i}/L_{\rm Edd}$, where $L_{\rm Edd}=1.26\times10^{38}(\mbh/\msun)\ \mathrm{erg\ s^{-1}}$; the on-state virial black hole mass is used for both epochs. This single-wavelength correction does not account for object-to-object differences in quasar SEDs, which can be characterized with multiband methods \citep[e.g.,][]{Chen2025Bolometric}. We therefore use the Eddington ratios only as secondary diagnostics.

\subsection{Matched Comparison Sample}\label{sec:matching}

The comparison sample is drawn from the fitting results for the extended quasars of \citet{Sun2026}. We apply the same quality cuts as for the CL AGNs and remove comparison quasars within 1\arcsec\ of a CL AGN, leaving 825 eligible comparison quasars. For the primary analysis, we match only the extended CL AGNs to extended DESI quasars with similar redshifts and host stellar masses. CL AGNs with compact morphology are not included in this comparison because their host properties are more sensitive to the degeneracy between the PSF and S\'ersic components. 

For each extended CL AGN, candidate comparison quasars satisfy $|\Delta z|\leq0.05$ and $|\Delta\log\mstar|\leq0.30$. We construct an optimal one-to-one match in the two-dimensional $z$--$\log\mstar$ plane without replacement, so that each comparison quasar is used once. Of the 88 extended CL AGNs, 82 can be paired with a unique comparison quasar within these limits. After matching, the mean redshift differs by 0.0006 and the mean $\log\mstar$ differs by 0.016 dex between the two samples.

For $\log\ssfr$ and $\hdelta_A$, we calculate the difference within each pair and derive the 95\% confidence interval on the mean difference from 200,000 paired bootstrap resamplings. Bootstrap resampling is not used for the quiescent and post-starburst fractions. For these fractions, we report Clopper--Pearson 95\% binomial confidence intervals \citep{ClopperPearson1934,1986ApJ...303..336G} and use two-sided McNemar tests to compare the matched pairs \citep{McNemar1947}. This calculation applies the binomial distribution to the pairs in which the CL AGN and comparison quasar have different classifications, and therefore accounts for the paired sample design. The quoted intervals describe sampling uncertainty and do not include the measurement uncertainties of individual decomposition quantities.

\section{Results}\label{sec:results}

\begin{figure}[t]
\centering
\includegraphics[width=\columnwidth]{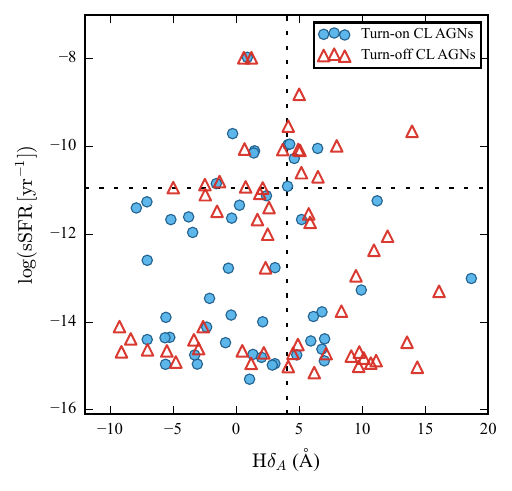}
\par\vspace{0.3em}
\includegraphics[width=\columnwidth]{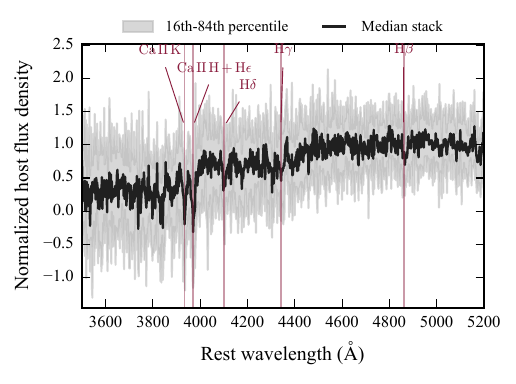}
\caption{Upper panel: $\hdelta_A$ measured from the decomposed DESI host spectrum versus sSFR. Blue filled circles and red open triangles denote turn-on and turn-off CL AGNs, respectively. Dashed lines show the adopted post-starburst criteria, and post-starburst host galaxies reside at the bottom right corner. Lower panel: median-stacked decomposed spectrum of the post-starburst hosts, displaying the Balmer absorption series and $\caii$ H and K.}
\label{fig:hdelta}
\end{figure}

\begin{figure*}[t]
\centering
\includegraphics[width=\textwidth]{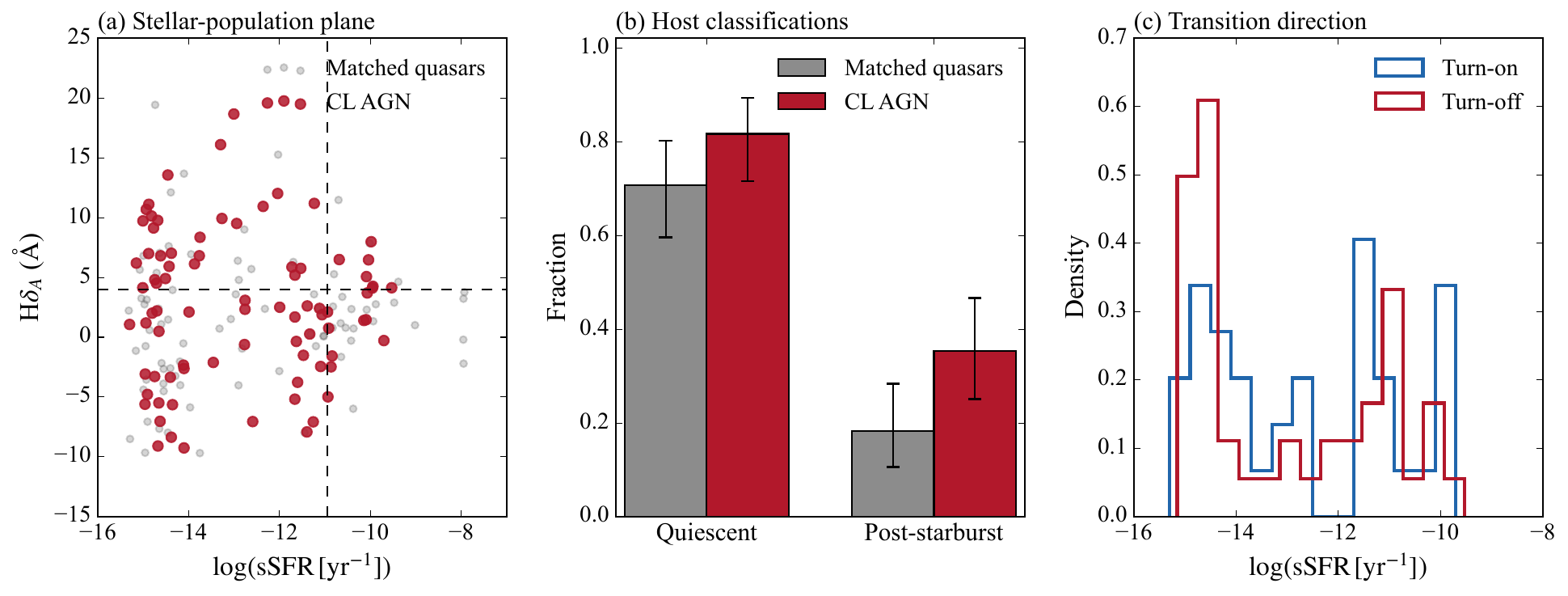}
\caption{Host galaxy comparison between extended CL AGNs and redshift--stellar-mass matched extended quasars analyzed with the same decomposition pipeline. The panels show the $\hdelta_A$--sSFR distribution, quiescent and post-starburst fractions, and state-separated sSFR distributions. Each CL AGN is paired with one unique comparison quasar. Error bars in the middle panel show 95\% binomial confidence intervals.}
\label{fig:matched}
\end{figure*}

\begin{table*}[t]
\centering
\caption{Primary matched sample comparison}
\label{tab:matched}
\begin{tabular}{lcccc}
\toprule
Outcome & CL AGNs & Matched quasars & CL AGNs $-$ quasars & Statistical result \\
\midrule
$\log(\ssfr/\mathrm{yr}^{-1})$ & $-12.87$ & $-12.62$ & $-0.25$ dex & 95\% interval $[-0.79,+0.28]$ \\
$\hdelta_A$ & $2.54$ \AA & $1.36$ \AA & $+1.18$ \AA & 95\% interval $[-0.59,+2.94]$ \\
Quiescent fraction & 67/82 (81.7\%) & 58/82 (70.7\%) & $+11.0$ percentage points & $p=0.108$ \\
Post-starburst fraction & 29/82 (35.4\%) & 15/82 (18.3\%) & $+17.1$ percentage points & $p=0.020$ \\
\bottomrule
\end{tabular}
\tablecomments{The first column lists the host property being compared. The second and third columns give the mean values of $\log\ssfr$ and $\hdelta_A$, or the numbers and fractions of classified objects, in the 82 unique matched pairs. The fourth column is the second column minus the third column. In the fifth column, $\log\ssfr$ and $\hdelta_A$ have 95\% confidence intervals from paired bootstrap resampling, whereas the quiescent and post-starburst comparisons have two-sided binomial probabilities calculated from pairs with different classifications. The 95\% binomial confidence intervals for the CL AGNs and matched quasars are 71.6--89.4\% and 59.6--80.3\%, respectively, for the quiescent fractions, and 25.1--46.7\% and 10.6--28.4\%, respectively, for the post-starburst fractions.}
\end{table*}

\subsection{Predominantly Quiescent Stellar Populations}

The host property analysis sample of 105 objects is dominated by low-sSFR hosts. We find that 79/105 (75.2\%) are quiescent according to their sSFR. Thirty-one objects (29.5\%) also have $\hdelta_A>4$ \AA\ and meet our post-starburst criterion. Figure~\ref{fig:hdelta} shows the distribution in the $\hdelta_A$--sSFR plane and a median stack of the post-starburst spectra. The Balmer absorption series is visible in both the stacked decomposed spectrum and the corresponding stellar models, supporting the post-starburst classifications.

The turn-off and turn-on subsets show similar host galaxy classifications. In the final sample of 105 objects, the post-starburst fractions are $35.7^{+7.5}_{-6.9}\%$ (20/56) for turn-off and $22.4^{+7.6}_{-6.3}\%$ (11/49) for turn-on systems, where the uncertainties are 68.27\% binomial confidence intervals. The two intervals overlap, indicating that the fractions are consistent within their individual $1\sigma$ binomial uncertainties; a two-sided Fisher test gives $p=0.198$. The corresponding quiescent fractions are 71.4\% (40/56) and 79.6\% (39/49), with $p=0.372$. Because state assignment depends on which survey observed the brighter epoch, and because the DESI spectra generally have lower S/N than the SDSS spectra, we do not interpret these numerical differences as evidence for distinct host galaxy populations among turn-on and turn-off CL AGNs.

\subsection{Comparison with Matched Extended Quasars}\label{sec:matched_results}

Figure~\ref{fig:matched} and Table~\ref{tab:matched} summarize the comparison between 82 extended CL AGNs and unique extended quasars at similar redshift and stellar mass. The CL AGNs have $\log\ssfr=-12.87$. Compared with $-12.62$ for the comparison quasars, the difference is $-0.25$ dex. The 95\% confidence interval, $-0.79$ to $+0.28$ dex, is consistent with no sSFR offset. For $\hdelta_A$, the corresponding values are $2.54$ and $1.36$ \AA, respectively. The difference is $+1.18$ \AA\ with a 95\% confidence interval of $-0.59$ to $+2.94$ \AA. Thus the extended CL AGNs do not show a clear offset from the comparison quasars in terms of either sSFR or $\hdelta_A$.

The quiescent fractions are 67/82 (81.7\%) for the CL AGNs and 58/82 (70.7\%) for the comparison quasars. The difference of 11.0\% is not statistically clear in the matched-pair test ($p=0.108$). The post-starburst fractions are 29/82 (35.4\%) and 15/82 (18.3\%), respectively. This difference of 17.1\% has a matched-pair probability of $p=0.020$.


\subsection{Extended Morphology and Host Detectability}\label{sec:morph_results}

\begin{figure*}[t]
\centering
\includegraphics[width=\textwidth]{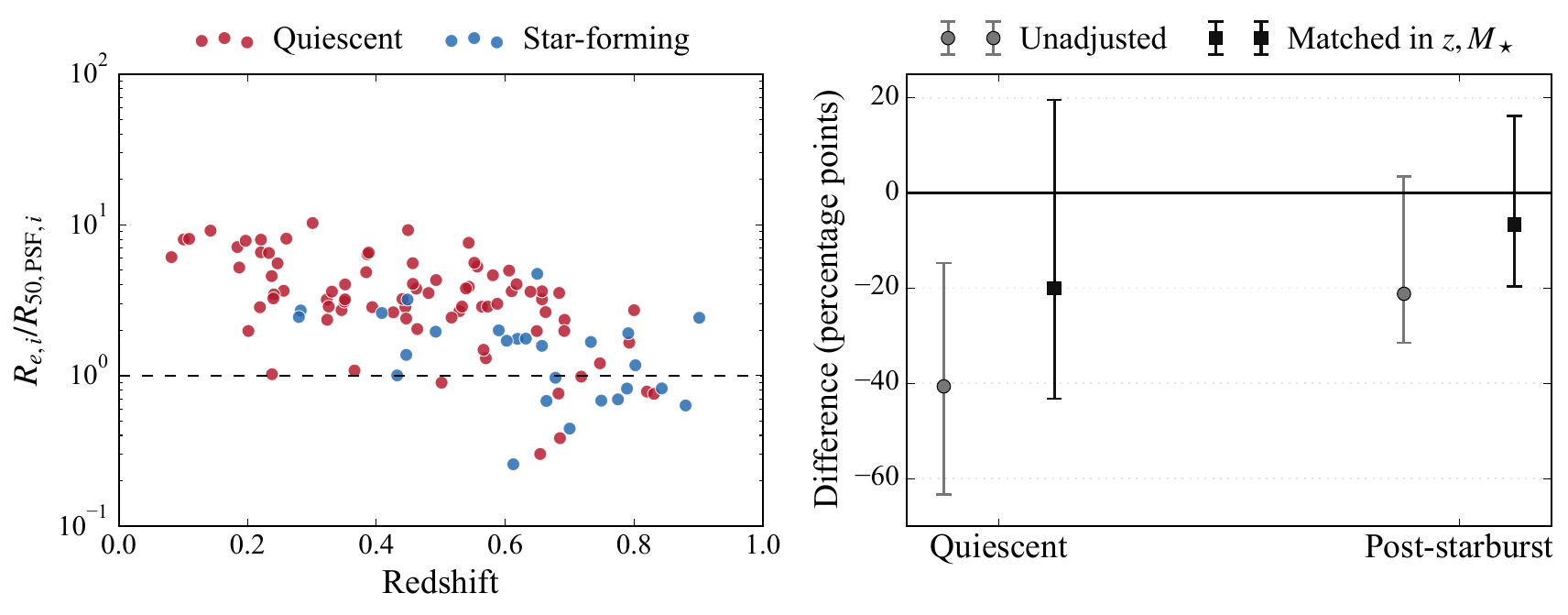}
\caption{Left: $i$-band S\'ersic effective radius relative to the HSC PSF half-light radius. Red and blue points denote quiescent and star-forming hosts, respectively; the dashed line marks the adopted threshold for extended morphology. Right: Observed differences in the quiescent and post-starburst fractions, defined as systems with compact morphology minus systems with extended morphology. Circles show the unadjusted comparison, and squares show the one-to-one comparison matched in redshift and stellar mass. Error bars show 95\% binomial intervals; the intervals for the matched comparison use the pairs with different classifications. The statistical probabilities are calculated with two-sided Fisher and McNemar tests, respectively.}
\label{fig:morphology}
\end{figure*}

\begin{table*}[t]
\centering
\caption{CL AGN host galaxy classifications by observed morphology}
\label{tab:morph_summary}
\begin{tabular}{lcccc}
\toprule
Subsample & $N$ & Median $z$ & Quiescent hosts & Post-starburst hosts \\
\midrule
CL AGN sample & 105 & 0.52 & 79/105 (75.2\%) & 31/105 (29.5\%) \\
Extended morphology & 88 & 0.46 & 72/88 (81.8\%) & 29/88 (33.0\%) \\
Compact morphology & 17 & 0.70 & 7/17 (41.2\%) & 2/17 (11.8\%) \\
Visually disturbed candidates & 8 & 0.47 & 7/8 (87.5\%) & 5/8 (62.5\%) \\
\bottomrule
\end{tabular}
\tablecomments{The first column lists the full sample and morphology subsamples. The second and third columns give the number of objects and median redshift. The fourth and fifth columns give the numbers and fractions of quiescent and post-starburst hosts. The 95\% binomial confidence intervals are 65.9--83.1\% and 21.0--39.2\% for the full-sample fractions, 72.2--89.2\% and 23.3--43.8\% for extended morphology, 18.4--67.1\% and 1.5--36.4\% for compact morphology, and 47.3--99.7\% and 24.5--91.5\% for the visually disturbed candidates.}
\end{table*}

Table~\ref{tab:morph_summary} summarizes the distribution for the final sample split by HSC morphology. The 88 CL AGNs with extended morphology are more often quiescent than the 17 systems with compact morphology (81.8\% and 41.2\%; two-sided Fisher $p=0.001$). The post-starburst fractions are 33.0\% and 11.8\%, respectively, but the difference is not statistically clear ($p=0.090$). These results are consistent with the extended quiescent host population found in \citet{Sun2026}. The median redshifts are 0.46 for the extended subsample and 0.70 for the compact subsample. The cosmic star formation rate density at these redshifts is $0.041$ and $0.060\ M_\odot\,\mathrm{yr}^{-1}\,\mathrm{Mpc}^{-3}$, respectively, following \citet{2014ARA&A..52..415M_cosmic_SFH}. As a simple estimate, scaling the 18.2\% star-forming fraction of the extended subsample by this factor of 1.48 gives 26.8\%, still well below the observed 58.8\% for the compact subsample. The same scaling corresponds to an sSFR increase of only 0.17 dex, whereas the mean $\log\ssfr$ values differ by 1.32 dex between the extended and compact subsamples ($-12.91$ and $-11.58$, respectively). Cosmic evolution alone therefore does not explain either difference.

\begin{figure*}[t]
\centering
\includegraphics[width=0.80\textwidth]{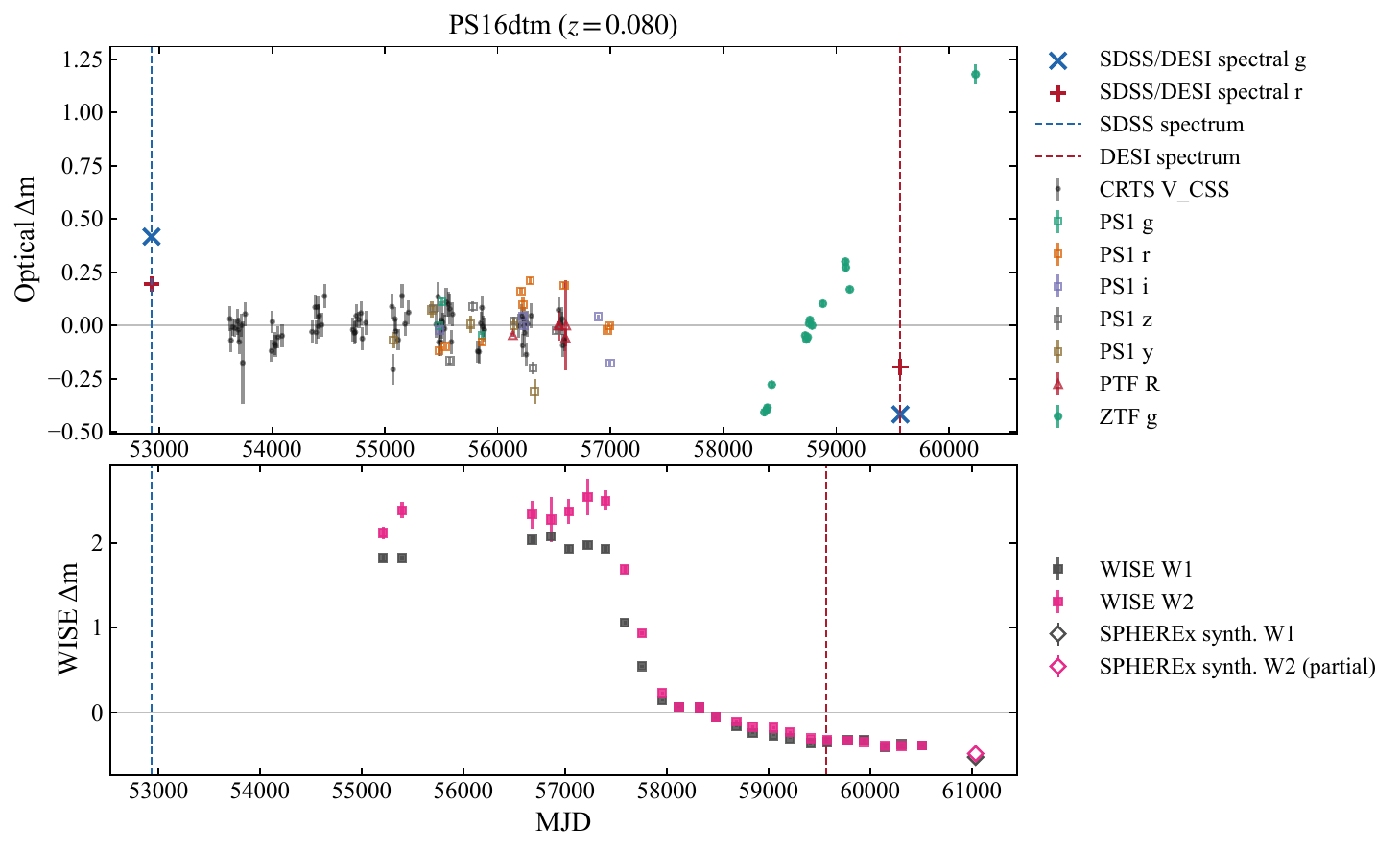}
\caption{Multi-survey differential light curves of ID 423 (PS16dtm), the previously identified TDE candidate among the DESI CL AGNs. Top: CRTS, PS1, PTF, and ZTF optical photometry. Bottom: WISE/NEOWISE W1 and W2. Open diamonds show W1 and W2 photometry synthesized by integrating the SPHEREx spectrophotometric spectrum over the covered WISE response curves; the W2 point is marked as partial because the available spectrum does not span the full W2 response. Each survey--filter series is centered on its own median; negative $\Delta m$ is brighter, and cross-survey offsets are not used. Dashed lines mark the SDSS and DESI spectral epochs; crosses show synthetic $g$- and $r$-band magnitudes from the SDSS and DESI spectra.}
\label{fig:ps16dtm}
\end{figure*}

For the primary morphology comparison, we pair each compact system with one extended CL AGN at similar redshift and stellar mass, using the same tolerances as in Section~\ref{sec:matching}. Fifteen compact systems have unique matches. The star-forming fractions are 60.0\% for the compact systems and 40.0\% for the matched extended systems. The difference of 20.0 percentage points is not statistically significant ($p=0.453$; Figure~\ref{fig:morphology}), and the paired sSFR values also show no clear offset. Thus, after controlling for redshift and stellar mass, the present sample does not establish a significant relation between host morphology and star formation. We use the subset with extended morphology for the external quasar comparison because it has the most reliable host decompositions and matches the selection criteria of the previous extended DESI quasar sample.

Eight visually disturbed candidates remain in the morphology sample of 105 objects. Five of the eight are classified as post-starburst, compared with 26/97 among the other objects (two-sided Fisher $p=0.047$). After one-to-one matching in redshift and stellar mass, the corresponding fractions are 5/8 and 1/8, with a McNemar probability of $p=0.125$. The sample is too small to establish a relation between disturbed structure and a post-starburst stellar population. A recent $z<0.15$ study reported a higher merger fraction among CL AGNs than in several comparison samples \citep{Tian2026Morphology}; the possible trend in our sample requires confirmation with a larger sample.

\section{Discussion}\label{sec:discussion}

\subsection{Long-Term Variability and TDE-Like Transients}\label{sec:lightcurve_results}

\begin{figure*}[t]
\centering
\includegraphics[width=\textwidth]{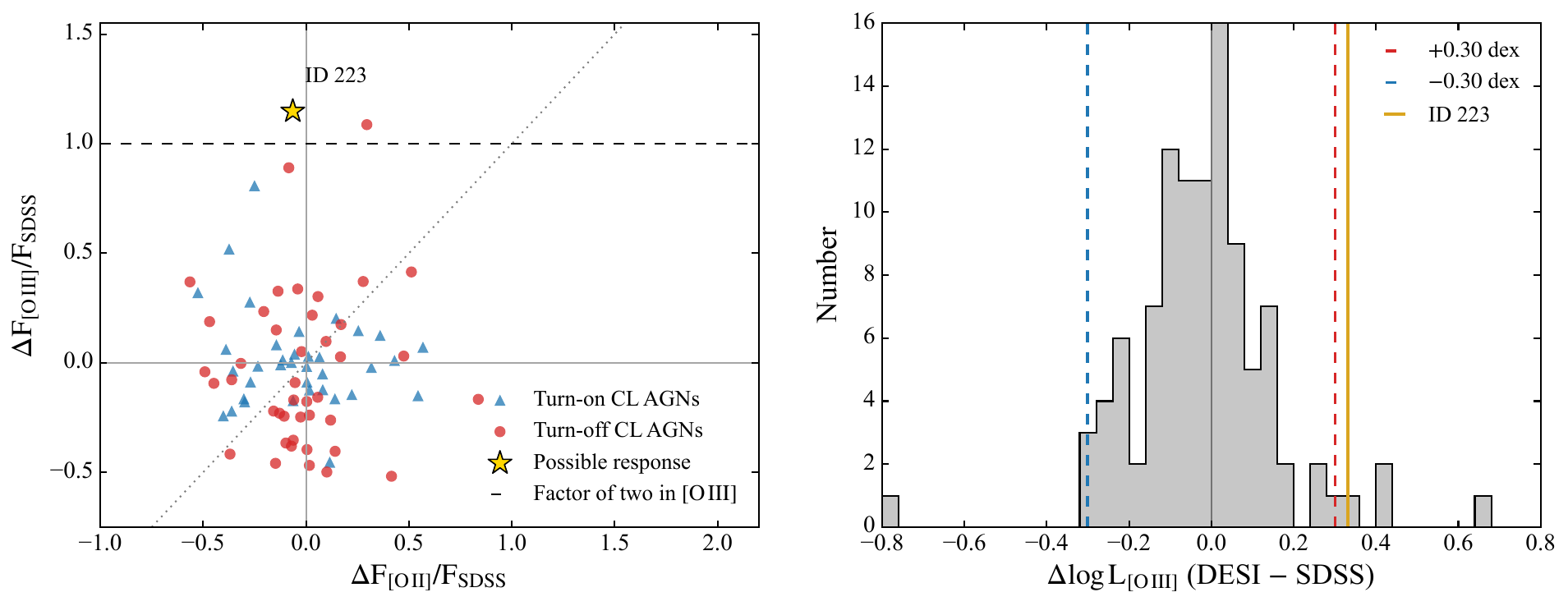}
\caption{Two-epoch forbidden line variability in the CL AGN sample. Left: fractional line-flux changes in $\oiii\ \lambda5007$ and $\oii\ \lambda3728$ for objects with reliable $\oii$ measurements, where $\Delta F\equiv F_{\mathrm{DESI}}-F_{\mathrm{SDSS}}$ and $F_{\mathrm{SDSS}}$ is the fitted line flux at the SDSS epoch. The star marks ID 223, a possible mild narrow line response. Right: distribution of $\Delta\log L_{\oiii}$. Positive values indicate a brighter DESI epoch, and dashed lines mark $\pm0.30$ dex.}
\label{fig:narrowline_variability}
\end{figure*}

PS1, WISE/NEOWISE, ZTF, and PTF yield measurements passing our quality cuts for 137, 137, 134, and 115 of the 137 HSC-matched CL AGNs, respectively. Usable CRTS photometry is available for two sources. The median optical baseline is 15.0 yr. The complete light curve atlas is provided as online supplementary material. Inspection of the light curves identifies nine systems with coherent, broad mid-infrared excursions that warrant further study.
Five of these nine have multiple optical brightening episodes in the same survey and filter, with the episode centers separated by more than one year. This indicates their long-term nuclear variability but does not distinguish recurrent accretion activity from delayed or repeating TDE emission. 
CL AGNs with IDs 264, 283, and 423 (PS16dtm) have at least two legacy optical surveys but no detected recurrence and remain consistent with a single event history at the present cadence; ID 113 also lacks detected recurrence but has only PS1 among the pre-ZTF surveys.

ID 423 (PS16dtm, Figure~\ref{fig:ps16dtm}) is the only previously identified TDE candidate in the DESI CL AGN subsample \citep{Guo2025DR1}. It is not included in our final host decomposition sample because its HSC image is PSF dominated, preventing a reliable host measurement. It nevertheless provides a useful case study of a TDE-like origin for CL behavior. PS16dtm has been studied extensively as a nuclear transient in a narrow-line Seyfert~1 galaxy. The original follow-up found a luminous optical and ultraviolet plateau, strong Fe~II emission, and X-ray suppression relative to the AGN state before the outburst, and interpreted the event as a TDE interacting with a pre-existing AGN disk \citep{Blanchard_2017}. Its mid-infrared flare was subsequently interpreted as a dust echo from circumnuclear material \citep{Jiang2017PS16dtmIR}. Monitoring over a longer baseline showed that the optical and ultraviolet transient faded while the mid-infrared emission remained bright for an unusually long time \mbox{\citep{Petrushevska2023PS16dtm,Jiang2025PS16dtmIR}}. In our light curve compilation, the W1 and W2 amplitudes are 2.42 and 2.88 mag (Figure~\ref{fig:ps16dtm}). The SPHEREx spectrophotometry adds an infrared spectral snapshot near Modified Julian Date (MJD) 61036; the synthetic W1 measurement and partial W2 measurement are consistent with the source remaining bright in the mid-infrared at that epoch. Taken together, these measurements are consistent with PS16dtm being an individual TDE-like event in an active nucleus. The other eight systems are suitable targets for follow-up observations, but their broad infrared excursions can also arise from changes in the accretion rate of an existing AGN. The present cadence is therefore insufficient for a TDE classification.

\subsection{Forbidden Narrow Line Stability}\label{sec:narrowline_results}

The BLR responds to changes in the ionizing continuum on timescales of days to months, whereas the more extended narrow line region (NLR) is expected to vary more slowly \citep{Bentz2013BLR,BaskinLaor2005OIII}. Narrow $\oiii$ is therefore commonly treated as constant during short AGN monitoring campaigns and is not expected to track rapid broad line transitions in CL AGNs \citep{vanGroningenWanders1992,PottsVillforth2021}. Compact NLR components can nevertheless vary on decadal timescales \citep{Peterson2013NLR}. We use the SDSS--DESI spectra to test for a coherent response of $\oiii$ and $\oii$ over multi-year baselines. Figure~\ref{fig:narrowline_variability} summarizes the two-epoch forbidden line measurements.

For each source, we define $\Delta\log L_{\mathrm{line}}\equiv\log L_{\mathrm{DESI}}-\log L_{\mathrm{SDSS}}$. In the left panel of Figure~\ref{fig:narrowline_variability}, the fractional line-flux change is $\Delta F_{\mathrm{line}}/F_{\mathrm{SDSS}}\equiv(F_{\mathrm{DESI}}-F_{\mathrm{SDSS}})/F_{\mathrm{SDSS}}$, where $F_{\mathrm{DESI}}$ and $F_{\mathrm{SDSS}}$ are the fitted line fluxes at the two spectral epochs. We use the total $\oiii\ \lambda5007$ luminosity, summing the fitted core and wing components. For $\oii\ \lambda3728$, we measure a local line flux from the decomposed AGN spectral component after subtraction of the tied host model.
The $\oiii$ variation analysis uses the 105 objects with reliable two-epoch decompositions, and the $\oii$ analysis uses the 80 of these objects with reliable local $\oii$ fits.

The DESI and SDSS spectra are separated by 2.36--18.39 yr in the rest frame, with a median baseline of 7.50 yr. Seven objects have an apparent $\oiii$ brightening larger than 0.30 dex, and three of them also have reliable $\oii$ measurements. Inspection of the corresponding spectral windows indicates that the two largest apparent changes, ID 177 and ID 497, are dominated by poor local spectral quality; we therefore exclude them from the population statistics and distributions in Figure~\ref{fig:narrowline_variability}. For the remaining 103 objects, the median $\Delta\log L_{\mathrm{line}}$ of $\oiii\ \lambda5007$ is $-0.017$ dex, with a median absolute deviation of 0.087 dex. Among the 79 objects with reliable local $\oii\ \lambda3728$ fits, the median change is $-0.018$ dex, with a median absolute deviation of 0.065 dex. Most objects remain close to zero change in both lines. ID 223 remains the only plausible individual narrow line response: $\oiii$ increases by 0.33 dex, while $\oii$ changes by only $-0.03$ dex.

The $\oiii$ luminosity change has a weak correlation with rest-frame baseline (Spearman $\rho=0.29$, $p=0.003$). A correlation of comparable strength is also present with redshift ($\rho=-0.33$, $p=8.1\times10^{-4}$), which is coupled to the physical fiber radius. Aperture and redshift effects therefore cannot be separated from a possible time dependence in this sample. The $\oii$ measurements with reliable local fits show no corresponding trends with baseline, redshift, or fiber size. The robust result is the absence of a population-wide $\oiii$ or $\oii$ response accompanying the changes in the continuum and broad lines. This stability provides population-level support for the standard spatial stratification of AGN line emission: the compact BLR responds rapidly, whereas the more extended NLR changes more slowly.

The lack of a population-wide response provides an approximate constraint on the size of NLR, $R_{\rm NLR}$. For a spherical emitting region, a change in the ionizing continuum is smoothed over the timescale $\tau_{\rm smooth}\approx 2R_{\rm NLR}/c+\tau_{\rm rec}$, where $\tau_{\rm rec}$ is the density-dependent recombination time \citep{BinetteRobinson1987,Peterson2013NLR}. If the measured narrow line emission were dominated by gas photoionized by the AGN, a CL transition occurred soon after the SDSS epoch, and $\tau_{\rm rec}$ were short, the median rest-frame baseline of 7.50 yr would imply $R_{\rm NLR}\gtrsim1.15$ pc; the longest baseline of 18.39 yr would give $R_{\rm NLR}\gtrsim2.82$ pc. These scales are comparable to the 1--3 pc $\oiii$ emitting region inferred from long-term variability in NGC 5548. They serve as estimated lower limits for the NLR inner edges of our CL AGNs because the exact transition time between the two states is unknown, and the measured line flux includes gas spanning a range of radii and densities. In low-density gas, the recombination time can itself be comparable to the epoch separation. Host galaxy emission further weakens the corresponding constraint from $\oii$.

\subsection{Black Hole Scaling Relations and Accretion State}

\begin{figure}[t]
\centering
\includegraphics[width=\columnwidth]{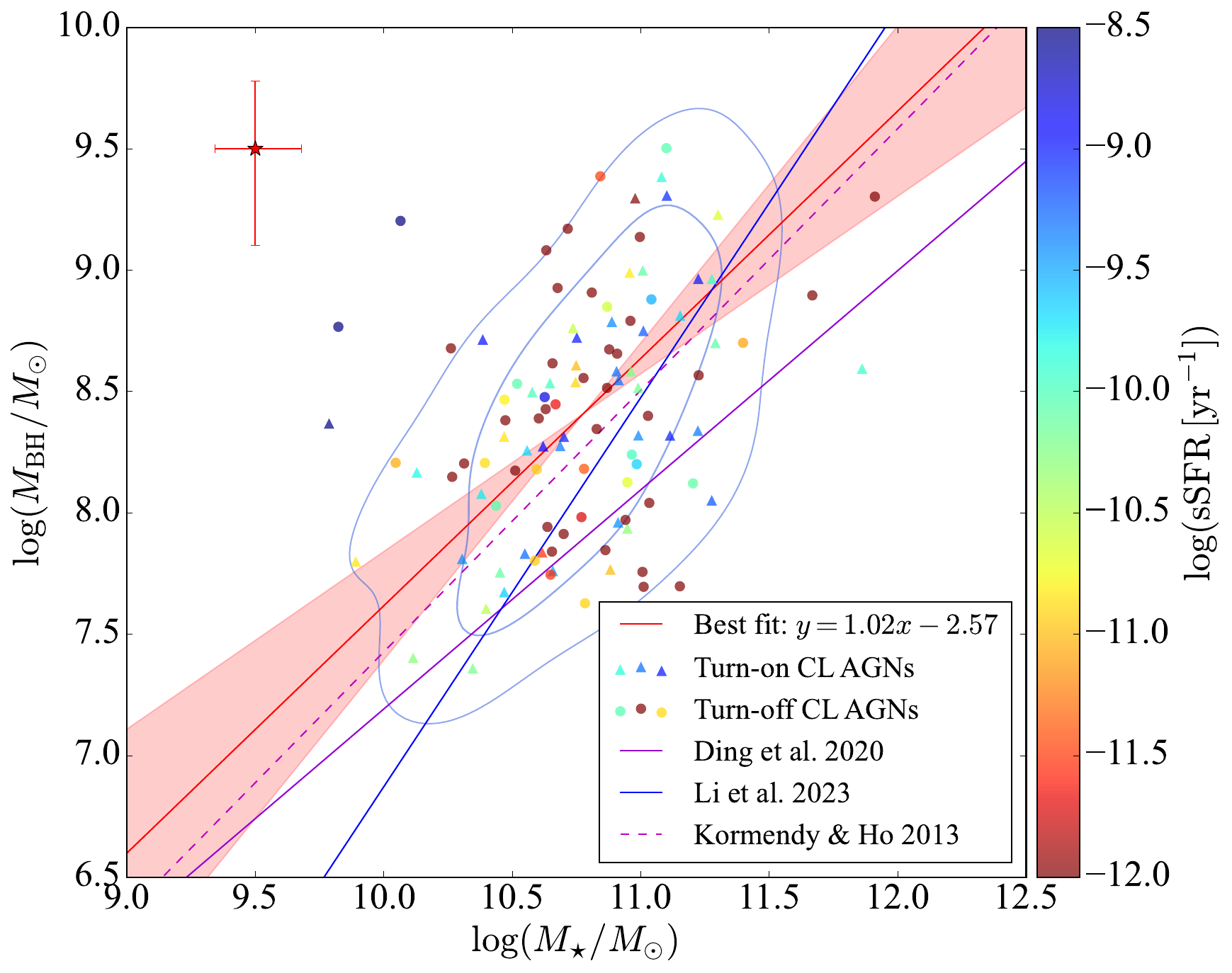}
\par\vspace{0.2em}
\includegraphics[width=\columnwidth]{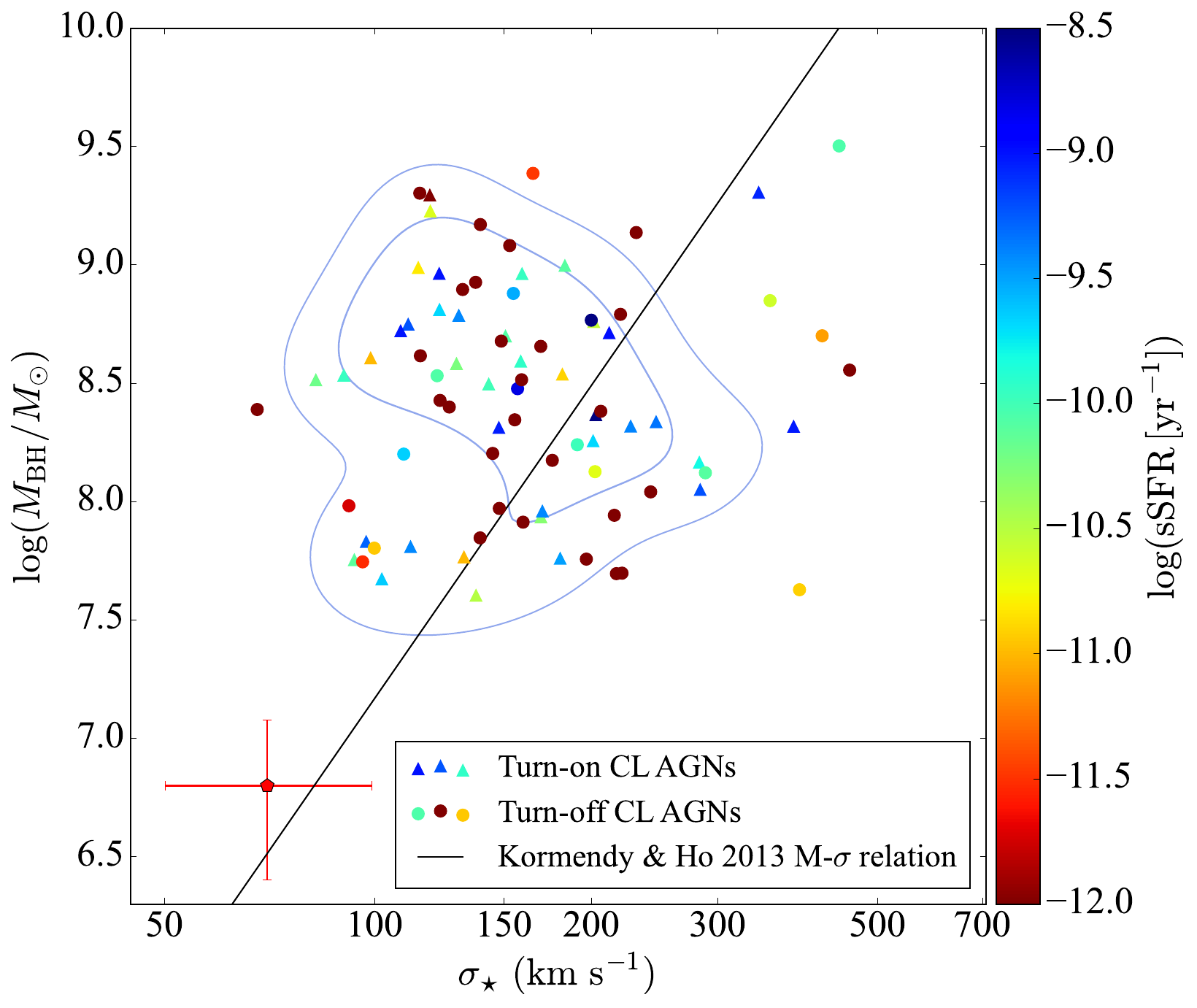}
\caption{Black hole mass versus host stellar mass (top) and stellar velocity dispersion (bottom). The red line in the top panel is an orthogonal distance regression fit to the 105 CL AGNs, with systematic uncertainties in $\log\mstar$ and $\log\mbh$ included in the fit. The bottom panel shows the 81 objects from the same sample with reliable host stellar velocity dispersion measurements. Blue contours indicate the two-dimensional distribution of the turn-on subset in each panel. The data do not constrain separate scaling relations for the turn-on and turn-off subsets.}
\label{fig:scaling}
\end{figure}

Figure~\ref{fig:scaling} places the CL AGNs on the $\mbh$--$\mstar$ and $\mbh$--$\sigma_{\star}$ planes. 
For reference, blue contours show the distribution of the turn-on subset.
For the $\mbh$--$\mstar$ plane, we fit the 105 CL AGNs in our final sample with an orthogonal distance regression accounting for the systematic uncertainties in $\log\mstar$ and $\log\mbh$, in addition to the statistical measurement uncertainties. The fitted slope is $1.02\pm0.14$, broadly consistent with the canonical local relations \citep[typically $\sim 1.1-1.2$; e.g.,][]{2013ARA&A..51..511K} and DESI quasars \citep{Sun2026} within the $1\sigma$ uncertainties.
We regard this slope as an empirical description because the sample spans a limited stellar mass range and the virial masses share common calibration systematics. The scatter around the fitted relation reflects uncertainties in single-epoch virial black hole masses, host stellar velocity dispersions measured from faint stellar components, kinematics integrated over the fiber aperture, and the narrow range of reliably recovered stellar masses. Apparent differences between the turn-on and turn-off distributions do not establish separate scaling relations. This result is consistent with previous CL AGN samples \citep{Jin2022,Yang2025,Zeltyn2026}.

The median Eddington ratios in the on and off states are $\log\lambda_{\mathrm{Edd}}=-1.30$ and $-1.49$, respectively. The modest state offset is qualitatively consistent with an accretion state transition near a few percent of Eddington \citep{NodaDone2018,RicciTrakhtenbrot2023}, but the absolute values inherit virial mass and bolometric correction systematics.

\subsection{Comparison with Other AGN Hosts}

The interpretation of the quiescent and post-starburst fractions depends on the comparison sample. Both the CL AGNs with extended morphology and the extended quasar comparison sample require detectable extended emission, successful HSC decomposition, and measurable stellar continua. This selection favors massive, low-sSFR systems with measurable host emission. Within the 82 matched pairs, the CL AGN quiescent fraction is higher by 11.0 percentage points. The CL AGN is quiescent while its comparison quasar is not in 17 pairs, whereas the reverse occurs in 8 pairs. The two-sided binomial probability for the paired classifications is $p=0.108$, equivalent to $1.6\sigma$, and does not show a statistically clear difference. For the post-starburst classification, the CL AGN alone is post-starburst in 23 pairs, compared with 9 pairs for the reverse. The 17.1 percentage point excess among the CL AGNs has $p=0.020$, equivalent to $2.3\sigma$, providing modest evidence for a higher post-starburst fraction.

The comparison is consistent with two strands of previous work. Hosts in the green valley and hosts with intermediate-age stellar populations are common in CL AGN samples \citep{2021ApJ...907L..21D,Jin2022}, whereas analyses with matched AGN comparison samples find broadly similar host star formation properties \citep{Verrico2025,Zeltyn2026}. Our analysis does not test whether post-starburst evolution contributes to black hole fueling. A post-starburst signature alone is not evidence that the changing-look event was triggered by a merger, a tidal disruption event, or rapid quenching across the galaxy.

\subsection{Separation of Galaxy and Nuclear Timescales}

The stellar diagnostics integrate over much longer timescales than a changing-look transition. Enhanced $\hdelta$ absorption traces A-star populations over hundreds of Myr, whereas the broad lines and continuum vary within years. A host can undergo multiple episodes of nuclear variability during one stellar population phase. The matched results do not establish a difference in the quiescent fraction, although they show a higher post-starburst fraction among the CL AGNs. This result does not imply a one-to-one association between a current CL event and quenching on galaxy scales.

The two-epoch decomposition turns this separation of timescales into a modeling constraint. The stellar population parameters are shared between DESI and SDSS, while separate aperture photometry accounts for the 1\farcs5 DESI fiber and the 2\arcsec\ or 3\arcsec\ SDSS fiber of each object. Consistent absorption features between epochs provide an internal check that complements the mock tests of \citet{Sun2026} and the tests based on repeated spectra in \citet{Aydar2026}.

Faint states that are nearly host-dominated provide internal validation of the decomposition, although they are too rare to define a separate population sample. We estimate the nuclear contribution in the off state from the fitted AGN component in representative continuum windows after masking the main broad line regions. Only two of the 105 objects have an off-state AGN contribution below 5\% of the observed continuum, and 15 are below 20\%. Both objects below 5\% are turn-on systems for which SDSS observed the faint state; among the 56 turn-off systems, for which DESI observed the faint state, the smallest off-state contribution is about 15\%. A single on-state spectrum would leave the host much more weakly constrained because the stellar continuum, AGN continuum, broad line wings, and flux calibration residuals can partially compensate for one another. Here, the faint-state spectrum and the HSC host photometry independently anchor the host flux level, while the bright-state spectrum constrains the variable nuclear component. ID 445 illustrates this point (Figure~\ref{fig:decomp_lowagn}): the SDSS faint state is nearly host-dominated and agrees with the host photometric constraints, whereas the DESI bright state requires a much stronger AGN component. The agreement between these two spectral states and the host photometry supports the physical consistency of the fitted host flux.

\subsection{Implications for the Changing-Look Mechanism}\label{sec:physical_origin}

The host galaxy analysis alone does not identify the mechanism responsible for a CL transition. The continuous stellar population measurements and quiescent classifications show no clear difference between the matched samples, whereas the post-starburst incidence is higher among the CL AGNs. The data do not establish that disturbed structure, a post-starburst episode, or quenching across the galaxy is synchronized with the spectral transition on timescales of years.

The nuclear measurements add constraints on the physical interpretation. Most objects in the early DESI sample become bluer when brighter \citep{2024ApJS..270...26G_CLAGN_from_EDR}. In the DR1 sample, broad line changes correlate with continuum changes, the Eddington ratio distribution turns over near $\lambda_{\mathrm{Edd}}\sim0.01$, and one slowly declining phase was interpreted as an accretion flow transition \citep{Guo2025DR1}. The difference in median Eddington ratio between our turn-on and turn-off states, together with the stable stellar component, is consistent with this picture. The combined evidence favors a change in the mass accretion rate or inner disk structure as the primary interpretation for CL behavior at the population level.

\begin{figure*}[t]
\centering
\includegraphics[width=\textwidth]{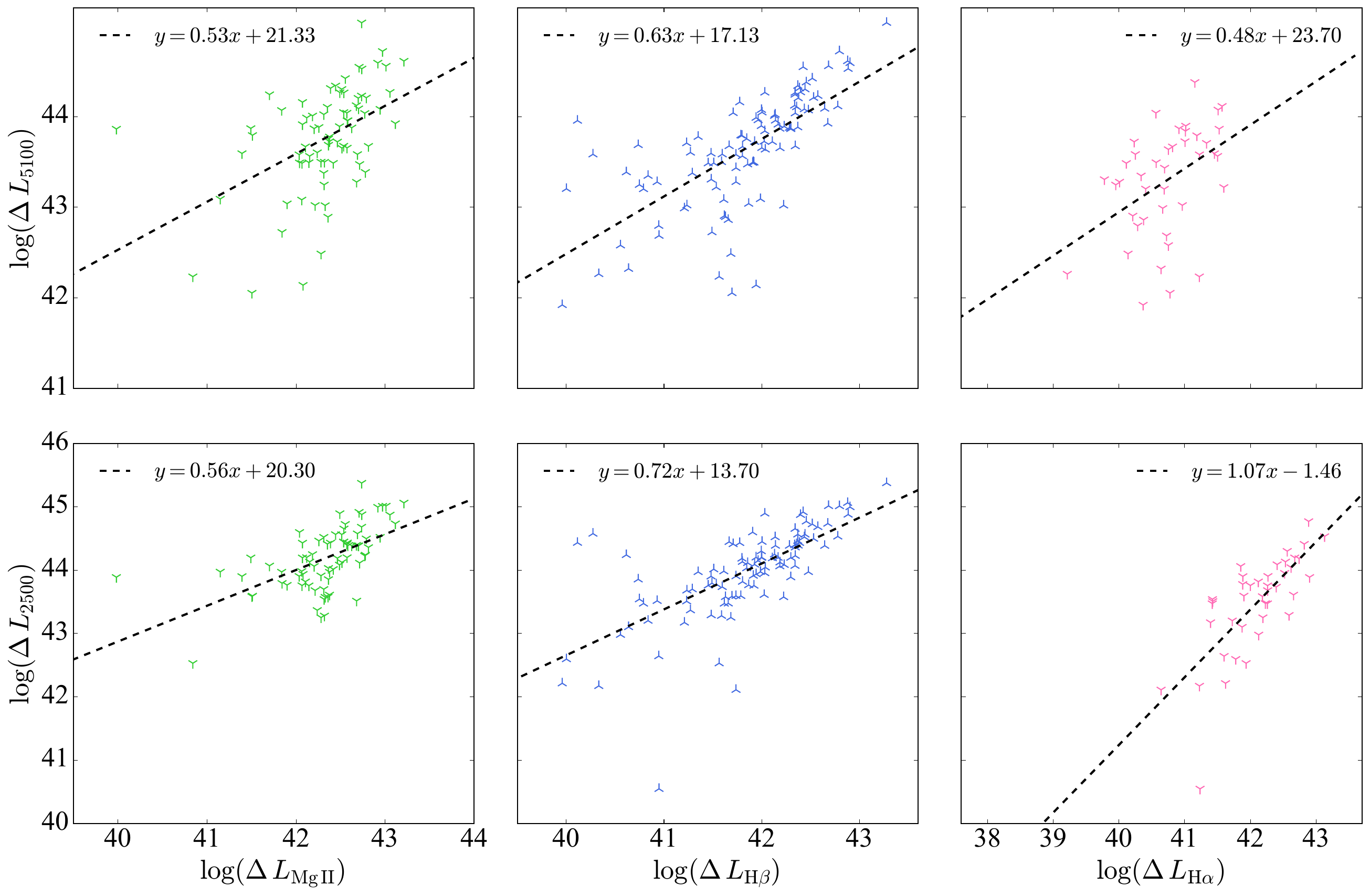}
\caption{Absolute changes in emission line luminosity versus the change in the decomposed AGN continuum luminosity between the DESI and SDSS epochs. Columns show $\mgii$, $\hb$, and $\ha$ from left to right. The top row uses $\lambda L_{\lambda}(5100\,\text{\AA})$, and the bottom row uses $\lambda L_{\lambda}(2500\,\text{\AA})$ with the same continuum definition. The positive correlations show that the broad line changes are coupled to nuclear continuum variability, while the different line responses are consistent with a stratified BLR.}
\label{fig:lvariation}
\end{figure*}

Figure~\ref{fig:lvariation} offers a consistency check using the decomposed AGN components of our CL AGN sample. In the top row, we compare the absolute change in the AGN component $\lambda L_{\lambda}(5100\,\text{\AA})$ between the DESI and SDSS epochs with the corresponding changes in $\mgii$, $\hb$, and $\ha$ line luminosity. The relations are positive in all three panels: the fitted slopes are 0.53, 0.63, and 0.48 for $\mgii$, $\hb$, and $\ha$, respectively, with $R^2=0.22$, 0.47, and 0.19. The bottom row shows the analogous comparison with $\lambda L_{\lambda}(2500\,\text{\AA})$ using the same continuum definition; the corresponding slopes are 0.56, 0.72, and 1.07. These positive correlations support a response of the broad lines to the varying nuclear continuum. The different line responses are also consistent with the stratified BLR behavior reported by \citet{Guo2025BLR}.

An accretion rate change need not explain every CL AGN. PS16dtm shows that a TDE-like flare can occur in a narrow-line Seyfert~1 galaxy and produce mid-infrared emission that remains bright for years because of circumnuclear dust \citep{Blanchard_2017,Jiang2017PS16dtmIR,Petrushevska2023PS16dtm,Jiang2025PS16dtmIR}.
Variable obscuration may also contribute to individual transitions, particularly in the absence of contemporaneous X-ray, polarization, and infrared constraints \citep{RicciTrakhtenbrot2023}. Confirmation of TDE and obscuration scenarios for individual sources requires further multiwavelength observations.

The forbidden line measurements provide an additional timescale constraint. The decomposed broad lines vary with the nuclear continuum, whereas the $\oiii$ and $\oii$ luminosities remain broadly stable for most objects over rest-frame baselines of several to nearly 20 yr. This contrast indicates that the observed transitions occur on spatial or temporal scales smaller than those probed by the gas dominating the forbidden line emission. It is consistent with an inner accretion flow change that rapidly alters the continuum and BLR while leaving the extended NLR and host galaxy environment approximately unchanged over the observed baselines.

\subsection{Selection Effects and Limitations}

Several limitations set the scope of our conclusions. First, the HSC coverage requirement selects 137 of 561 parent CL AGNs, and successful decomposition further requires a measurable host contribution and an acceptable HSC imaging decomposition. The reported fractions apply to CL AGNs with measurable host emission, not to the whole population. Second, SFRs near the boundaries of the nonparametric SFH model are effectively upper limits; their numerical values should not be interpreted as measurements of extremely small SFR. 

Third, $\hdelta_A$ is sensitive to flux calibration, AGN subtraction, and spectral S/N. Applying the same measurement to the comparison sample mitigates, but does not eliminate, these systematics. Fourth, matching to type~1 extended quasars cannot fully reproduce the dependence of CL AGN detectability on spectral state. Fifth, the morphology split is based on a single size metric from ground-based imaging, while the disturbed classification is visual and contains only eight objects with reliable host properties. Finally, neither comparison accounts for environment or fiber covering fraction. Larger samples with uniform imaging at higher spatial resolution, repeated spectroscopy, and explicit structural matching will improve these analyses.

The current data do not support two further inferences. Stellar metallicity derived from joint broadband photometry and optical spectroscopy does not measure the variable nuclear obscuring column and therefore cannot rule out changing obscuration. Likewise, the narrow line measurements do not separate extended NLR emission from host emission powered by star formation, because the fibers differ in size and the emitting gas is spatially extended and mixed.

\section{Summary}\label{sec:summary}

We applied a two-epoch extension of the spectrophotometric decomposition developed by \citet{Sun2026} to SDSS and DESI spectra of 105 CL AGNs with HSC imaging. The model uses a common host stellar population across the two epochs, while allowing for aperture-dependent host contributions and independently varying AGN components. Our main results are as follows.

\begin{enumerate}
\item The host sample is predominantly quiescent: 79/105 (75.2\%) objects have $\log\ssfr<-10.94$, and 31/105 (29.5\%) also satisfy the post-starburst criterion of $\hdelta_A>4$ \AA. The turn-on and turn-off subsets do not show statistically significant differences in either classification.

\item For 82 one-to-one pairs matched in redshift and stellar mass, CL AGNs with extended host emission and extended quasars have consistent $\log\ssfr$, $\hdelta_A$, and quiescent fractions. The post-starburst fraction is higher among the CL AGNs, with modest statistical evidence for a difference ($p=0.020$).

\item In our CL AGN sample, systems with compact morphology are more often star-forming than those with extended host emission. The redshift dependence of the cosmic star formation rate is too small to explain the observed difference. However, the difference is not statistically significant after matching in redshift and stellar mass, and host measurements for compact systems are more sensitive to nuclear contamination. The present sample therefore does not establish an intrinsic relation between morphology and star formation.

\item The decomposed AGN continuum variations correlate with the changes in broad $\mgii$, $\hb$, and $\ha$ emission. In contrast, $\oiii$ and $\oii$ show no population-wide response over the SDSS--DESI baselines. This difference is consistent with a compact BLR that responds rapidly to the ionizing continuum and a more extended NLR that varies more slowly.

\item The black hole mass--stellar mass relation is consistent within the uncertainties with local relations and that of extended DESI quasars. The turn-on and turn-off systems do not define separate scaling relations. The multi-survey light curves identify individual events that warrant further study, including PS16dtm, but do not establish a TDE origin for the CL AGN population.
\end{enumerate}

At the population level, CL AGNs with measurable extended host emission do not occupy a distinct host galaxy population from similarly selected extended quasars. Together with the correlated continuum and broad line variations, this result favors changes in the central black hole accretion rate or inner disk structure as the main origin of CL transitions, while TDEs and variable obscuration remain possible explanations for individual events.

\begin{acknowledgments}

We acknowledge the science research grants from the China Manned Space Project (CMS-CSST-2021-A05) and the National Science Foundation of China (12225301). SEIB is supported by the Deutsche Forschungsgemeinschaft (DFG) under Emmy Noether grant number BO 5771/1-1a.

This research used data obtained with the Dark Energy Spectroscopic Instrument (DESI). DESI construction and operations is managed by the Lawrence Berkeley National Laboratory. This material is based upon work supported by the U.S. Department of Energy, Office of Science, Office of High Energy Physics, under Contract No. DE-AC02-05CH11231, and by the National Energy Research Scientific Computing Center, a DOE Office of Science User Facility under the same contract. Additional support for DESI was provided by the U.S. National Science Foundation, Division of Astronomical Sciences under Contract No. AST-0950945 to the NSF's National Optical-Infrared Astronomy Research Laboratory; the Science and Technology Facilities Council of the United Kingdom; the Gordon and Betty Moore Foundation; the Heising-Simons Foundation; the French Alternative Energies and Atomic Energy Commission (CEA); the National Council of Humanities, Science and Technology of Mexico (CONAHCYT); the Ministry of Science and Innovation of Spain (MICINN); and the DESI Member Institutions: \url{https://www.desi.lbl.gov/collaborating-institutions}. The DESI Collaboration is honored to be permitted to conduct scientific research on I'oligam Du'ag (Kitt Peak), a mountain with particular significance to the Tohono O'odham Nation. Any opinions, findings, and conclusions or recommendations expressed in this material are those of the authors and do not necessarily reflect the views of the U.S. National Science Foundation, the U.S. Department of Energy, or any of the listed funding agencies.

Funding for the Sloan Digital Sky Survey IV was provided by the Alfred P. Sloan Foundation, the U.S. Department of Energy Office of Science, and the participating institutions. SDSS-IV acknowledges support and resources from the Center for High-Performance Computing at the University of Utah.

This work is based on data from the Hyper Suprime-Cam Subaru Strategic Program collected at the Subaru Telescope and retrieved from the HSC data archive operated by the Subaru Telescope and Astronomy Data Center at the National Astronomical Observatory of Japan. We acknowledge the HSC collaboration and the teams that developed the HSC instrumentation and software. We are honored and grateful for the opportunity to observe the Universe from Maunakea, which has cultural, historical, and natural significance in Hawaii.

We acknowledge the survey teams that produced the public CRTS, PS1, PTF, and ZTF photometry used in this work. CRTS is supported by NASA and the U.S. National Science Foundation, and ZTF by the U.S. National Science Foundation and its partner institutions. The PS1 data were obtained from the Mikulski Archive for Space Telescopes at the Space Telescope Science Institute; the PTF and ZTF data were obtained from the NASA/IPAC Infrared Science Archive.

This work also uses data products from the NASA WISE, NEOWISE, and SPHEREx missions. This research has made use of the NASA/IPAC Infrared Science Archive, which is funded by NASA and operated by the California Institute of Technology.

The archival data products are \dataset[PS1 DR2]{https://doi.org/10.17909/s0zg-jx37}, \dataset[PTF light curves]{https://doi.org/10.26131/IRSA647}, \dataset[WISE single exposure photometry]{https://doi.org/10.26131/IRSA139}, \dataset[NEOWISE single exposure photometry]{https://doi.org/10.26131/IRSA144}, \dataset[ZTF light curves]{https://doi.org/10.26131/IRSA598}, and \dataset[SPHEREx QR2 spectral images]{https://doi.org/10.26131/IRSA652}.

\end{acknowledgments}

\facilities{DESI, SDSS, Subaru (HSC), CRTS, MAST (Pan-STARRS), PTF, WISE, NEOWISE, ZTF, SPHEREx, IRSA}
\software{Astropy \citep{Astropy2013,Astropy2022}, \texttt{Bagpipes} \citep{10.1093/mnras/sty2169}, \texttt{GalfitM} \citep{2022A&A...664A..92H_galfitm}, Matplotlib \citep{Hunter2007}, NumPy \citep{Harris2020}, pandas \citep{McKinney2010}, \texttt{PyQSOFit} \citep{2018ascl.soft09008G}, SciPy \citep{Virtanen2020}}

\appendix
\setcounter{figure}{0}
\renewcommand{\thefigure}{A\arabic{figure}}
\renewcommand{\theHfigure}{A\arabic{figure}}
\section{Narrow Line Diagnostic Diagrams}\label{app:bpt}

\begin{figure*}[t]
\centering
\includegraphics[width=0.49\textwidth]{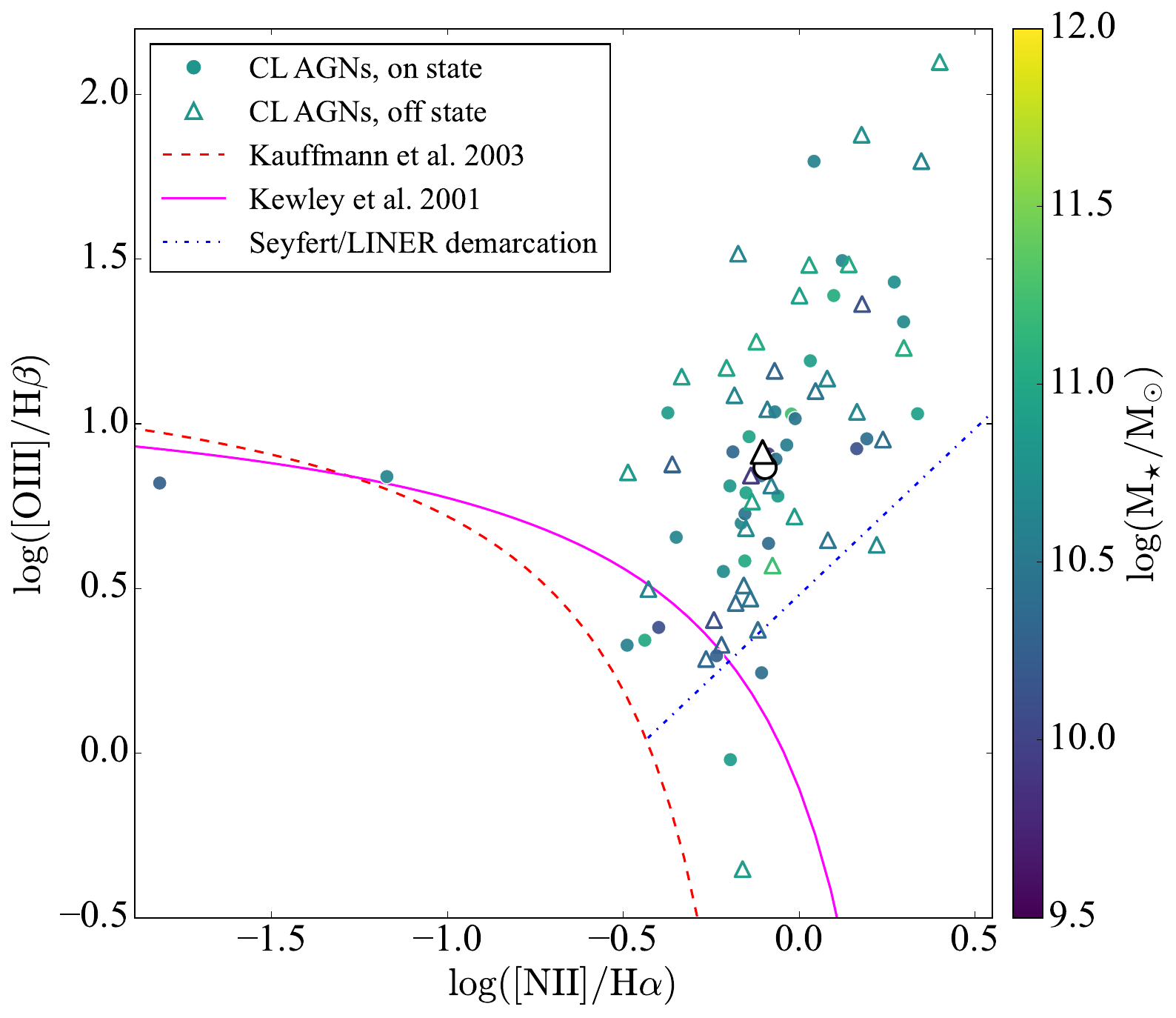}
\includegraphics[width=0.49\textwidth]{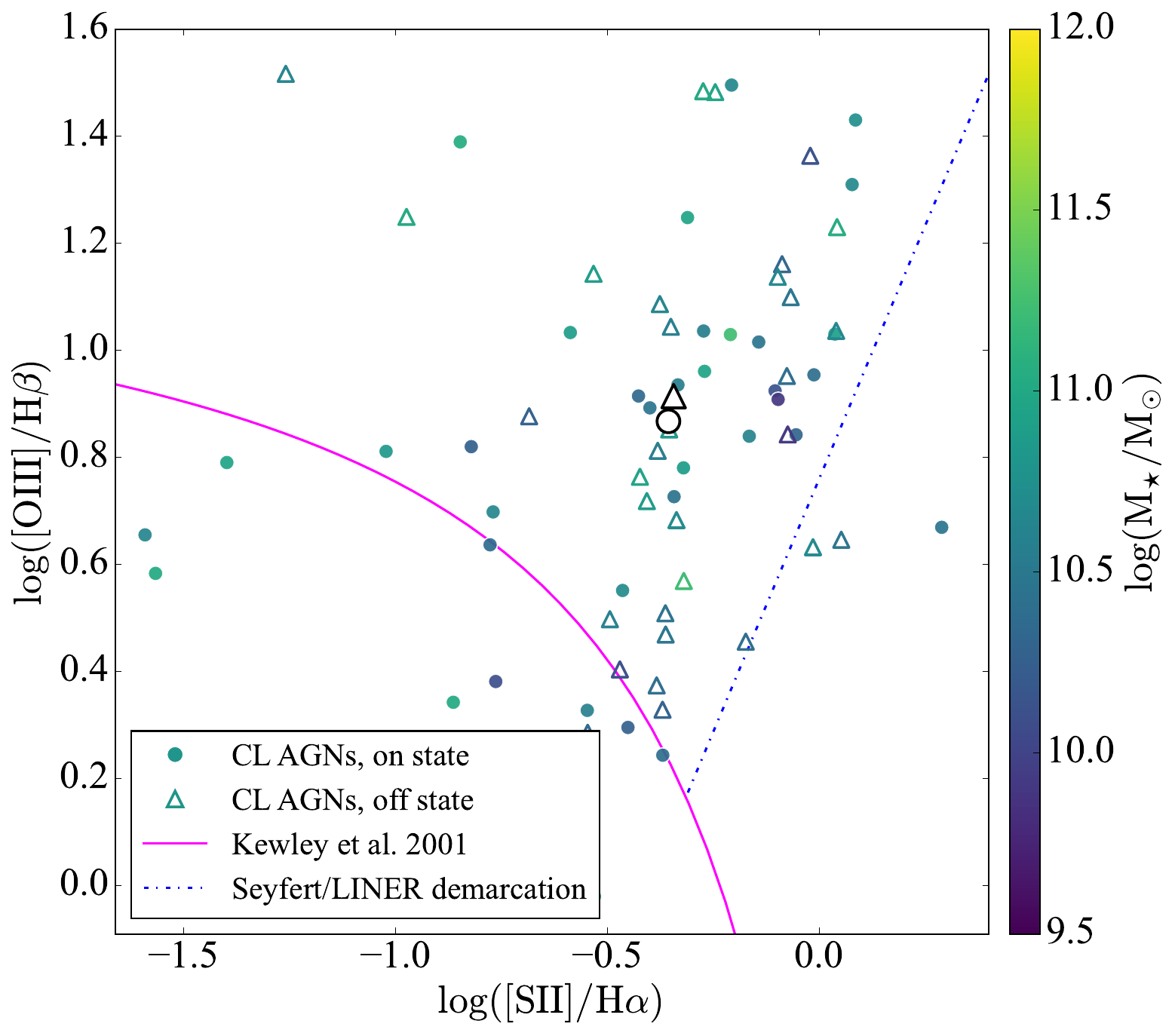}
\caption{Narrow line diagnostic diagrams for the decomposed DESI and SDSS spectra. The points are color-coded by stellar mass; filled circles and open triangles show the on and off states, respectively. Enlarged black-edged symbols mark the sample medians. Different fiber sizes and the longer response timescale of the NLR limit a direct interpretation of the median offsets between states.}
\label{fig:bpt}
\end{figure*}

Figure~\ref{fig:bpt} shows the Baldwin, Phillips \& Terlevich (BPT) diagnostic diagrams \citep{Baldwin1981}, comparing $\oiii/\hb$ versus $\nii/\ha$ and $\sii/\ha$. We use the theoretical maximum starburst boundaries of \citet{Kewley2001} in both panels and the empirical \citet{Kauffmann2003} boundary in the $\nii$ panel. The Seyfert--low-ionization nuclear emission-line region (LINER) divisions are the BPT plane transposition of \citet{CidFernandes2010} in the $\nii$ panel and the empirical boundary of \citet{Kewley2006} in the $\sii$ panel. The slight median displacement between on and off states is smaller than the combined aperture and decomposition systematics. We report these diagrams as a consistency check on the narrow line measurements and do not use the median offsets to infer a change in the dominant ionization source.

\bibliography{clq_ref}
\bibliographystyle{aasjournalv7}

\end{document}